# Synthesized Kuramoto potential via optomechanical Floquet engineering

Motoki Asano[1]*, Hajime Okamoto[1], and Hiroshi Yamaguchi[1]

[1]NTT Basic Research Laboratories, NTT Corporation, 3-1 Morinosato Wakamiya, Atsugi-shi, Kanagawa 243-0198, Japan

*Corresponding author. Email: motoki.asano@ntt.com

**Abstract:** Synchronization is a ubiquitous scientific phenomenon in various physical systems. Here, we examine the feasibility of generating multistable and dynamically tunable synchronization by using the technique of Floquet engineering. Applying a periodically modulated laser light to optomechanical oscillators allows for stable and precise control of oscillator couplings. This enables us not only to explore the physics of quantized integer and fractional phase slips but also synthesize multioctave synchronizations of mechanical oscillators that exhibit tailorable multistability. Furthermore, the dynamically manipulated synchronizations lead to an exotic topology wherein the phase trajectories have a nontrivial winding number and giant non-reciprocity. This scheme could help to elucidate the dynamics of complicated oscillator networks like biological systems and to mimic their highly efficient information processing.

Synchronization is a universal phenomenon appearing in not only macroscopic systems such as fire-fly fluorescence (1) and heart pulsation (2), but also nanoscale quantum worlds such as atomic clusters (3) and superconducting circuits (4). In particular, synchronization in oscillator networks has been exploited by innovative technologies such as secret communications (5,6), non-von Neumann computation (7), and efficient power grids (8,9). It is thought that implementing engineered synchronization in nano- or mesoscale integrated devices, such as mechanical resonators (10-15), lasers (16,17), and superconducting circuits (4,18,19), will pave the way for efficient device-based information processing mimicking the functional processes found in macroscopic systems.

Nonlinear synchronization dynamics in mesoscale oscillator devices can be comprehensively described by the Kuramoto model (20,21). For $N$ oscillators, the standard Kuramoto model represents the dynamics of the oscillation phase, $\varphi_j$, as $\dot{\varphi}_j = \omega_j + \frac{K}{N}\sum_{j=1}^{N}\sin(\varphi_j - \varphi_i)$ where $\omega_j$ denotes the $j$th oscillator frequency and $K$ denotes the coupling strength. For the case of two coupled oscillators ($N = 2$), this equation can be reduced to a single differential equation of the phase difference $\varphi_{\text{diff}} \equiv \varphi_1 - \varphi_2$ governed by a simple sinusoidal potential, $\dot{\varphi}_{\text{diff}} = \Delta + K\sin(\varphi_{\text{diff}})$, with $\Delta = \omega_1 - \omega_2$. The standard model can be modified by introducing an additional nonlinear potential with a different shape (here, we will also refer it as the Kuramoto potential). This modification successfully predicts phenomena such as synaptic plasticity (22), chaos synchronization (5,6), and chimera state formation (23). However, the above phenomena are hard to exhibit in conventional devices whose designs seek to limit nonlinearities in their operation and in which the shape of the Kuramoto potential can be controlled only very limitedly.

Here, we demonstrate tailor-made and dynamically controllable phase synchronization by artificially synthesizing a Kuramoto potential with a cavity optomechanical setup. Cavity optomechanics, where cavity photons interact with mechanical vibrations via radiation pressure, provides a highly controllable means of generating mechanical nonlinearity through the use of a high-Q microresonator (24). Because the optomechanical coupling strength depends on the number of cavity photons, intensity modulation of the input laser light allows for time-domain control of its nonlinearity, which is known as optomechanical Floquet engineering (25,26).

We used this optomechanical Floquet engineering approach to tailor a two-mode Kuramoto potential based on mixed-overtone injection and control it adiabatically. This precise and functional scheme enables us to manipulate not only integer but also fractional phase slips, which is a key to inducing multistability in phase synchronization. Furthermore, it lets us examine as yet unexplored functions in synchronization: asymmetric bistability and topological non-reciprocal trajectories. Thus, optomechanical Floquet engineering could significantly expand the functionality of synchronization phenomena appearing in solid-state oscillator networks.

**Results**

**Two-mode phase synchronization in cavity optomechanics**

Describing the system dynamics using a phase-space potential provides us with a good illustration of the idea of synthesized synchronization (see also the Supplementary Materials). Here, we consider a modification in which mixed overtones are included in the Kuramoto potential, $\zeta(\varphi)$, defined as

$$\dot{\varphi}_{\text{diff}} = -\frac{\partial \zeta}{\partial \varphi_{\text{diff}}}, \qquad \zeta = \Delta \varphi_{\text{diff}} + \sum_{l=1}^{\infty} A_l \cos(l\varphi_{\text{diff}} + \theta_l). \tag{1}$$

The second term is the mixed-overtone potential, which leads to there being equilibrium points in phase space through superposition of sinusoidal functions with control parameters $A_l$ and $\theta_l$. For instance, the simplest synchronization, denoted by $\zeta_0 = A \cos\varphi_{\text{diff}}$ ($\Delta = 0, A_{l\neq 1} = 0, \theta_l = 0$), corresponds to the standard Kuramoto model for two coupled identical oscillators and shows a monostable equilibrium point at $\varphi_{\text{diff}} = \pi$. This phenomenon is analogous to what is observed in a Josephson junction, where the sinusoidal washboard potential maintains the quantum phase correlation across the junction. Moreover, in this study, we demonstrated arbitrary control of this potential by superposing the first- and a higher order sinusoidal potential ($l \geq 2$) and adiabatically modifying $A_l$ and $\theta_l$.

To make this idea operational, we focus on a cavity optomechanical setup that consists of two-mechanical modes with different frequencies simultaneously coupling to optical modes via radiation pressure (see Fig. 1A). Because the optomechanical interaction is dispersive, cavity photons play a role of a nonlinear medium for the two mechanical modes instead of the standard structural mechanical nonlinearities (27,28). This nonlinearity in the mechanical modes, especially the third-order Duffing nonlinearity, induces a periodic Kuramoto potential as it does in nanomechanical systems (12). Furthermore, this nonlinearity can be tuned by modulating the cavity photon number, i.e., input laser intensity. In this work, periodically modulating the cavity photon number, i.e., optomechanical Floquet engineering (25,26), was utilized for inducing phase synchronization. Modulation (at $\Omega_{\text{mod}}$) around the difference in frequency ($\Omega_{\text{diff}}$) between the two mechanical modes induces locking of their phase difference to the optical modulation phase $\theta_{\text{mod}}$. This scheme of non-degenerate synchronization is theoretically described in detail in the Supplementary Materials, where the frequency detuning is given by $\Delta = \Omega_{\text{mod}} - \Omega_{\text{diff}}$.

In our experiment, a single microbottle resonator (29,30) fabricated on a silica glass fiber was used as a cavity optomechanical platform with high-Q optical whispering gallery modes (WGMs) simultaneously coupling to multiple mechanical radial breathing modes (RBMs) via radiation pressure (see Fig. 1B). An optical Q factor of about $10^6$, a mechanical Q factor of about $10^3$, and their optomechanical coupling constant of about $10^2$ Hz were achieved similarly to in conventional microbottle resonators (31,32). As a prominent feature of the microbottle resonator, multiple RBMs appear with different axial mode numbers depending on the curvature of the bottle. Our device has low-finesse optical spectral profiles which cause simultaneous spontaneous oscillation of two mechanical modes (33). The simultaneous oscillation spectra were obtained around 48 MHz when 20-mW telecommunication laser light was coupled to the WGM (see Fig. 1C). The detailed experimental setup is shown in the Supplementary Materials.

To form a Kuramoto potential between two oscillating RBMs via optomechanical Floquet engineering, the intensity of the input laser light was modulated around the difference frequency of two mechanical oscillators ($\Omega_{\text{mod}} \sim \Omega_{\text{diff}} = 2\pi \cdot 470$ kHz). The detuning $\Delta$ and modulation depth ($A_l$), i.e., the voltage applied in the optical intensity modulator $V_{\text{mod}}$, determines the tilt and depth of the potential, respectively (see Fig. 1D). Because $\Omega_{\text{mod}}$ (tilt) and $V_{\text{mod}}$ (depth) are optically tunable, we can explore the synchronization condition by sweeping $V_{\text{mod}}$ for each $\Omega_{\text{mod}}$. This technique, referred to as adiabatic amplitude modulation, enables us to promptly confirm where the synchronization is stabilized in the parameter space and it avoids having to contributions from long-term noise (e.g., laser drift). Figure 1E shows a color map of measured amplitudes of the two-mode beat notes. The higher beat-note amplitude reflects phase synchronization where the relative phase of two-mode oscillations is locked. In fact, the synchronization area has a triangular shape (a.k.a. an Arnold tongue) that is typical of phase synchronization. Moreover, higher-order synchronization occurs at integer multiples of the modulation frequency. Triangular structures were also observed for these higher synchronization orders (Supplementary Materials).

Besides the beat-note measurement, the stability of the frequency difference between the two modes also confirmed that phase synchronization had occurred. The fluctuation of the frequency difference under synchronization was remarkably suppressed compared with the unsynchronized case (see Fig. 1F). Figure 1G depicts a quantitative stability analysis in terms of the Allan deviation for the three different synchronization orders $n = 1, 2$, and 3. Obviously, the long-term stability ($\tau_{\text{int}} > 0.3$ sec) was dramatically improved under synchronization compared with the unsynchronized case.

**Quantized phase slip at the edge of synchronization regime**

Because our approach provides well-engineered synchronization conditions, we can explore phenomena at the boundary between the synchronous and asynchronous regimes. The outer neighborhood of the domain edge shows a periodic transition between stable and unstable states where the potential depth ($A_l$) is slightly smaller than the tilt ($\Delta$). This edge of the synchronization regime causes the quantized phase slip along the tilted potential (see Fig. 2A). This phenomenon is analogous to what occurs in Josephson junctions in superconducting circuits, where the applied voltage tilts the washboard potential and discontinuously shifts the phase difference around its critical current (34,35). In the case of $n = 1$, the phase slip is quantized as $\delta\phi_{\text{slip}} = 2\pi$. By precisely controlling the detuning, we observed a $2\pi$ phase slip with different slip durations (see Fig. 2B). With increasing potential tilt i.e., $\Delta$, the phase slip duration, $\tau$, becomes shorter because the translational motion along the direction of tilt becomes fast enough for the oscillations to escape the potential minima. Moreover, since the higher order synchronization makes multistability in the phase difference, the amount of quantized phase slip depends on the synchronization order. Phase quantizations with $\delta\phi_{\text{slip}}(n) = 2\pi/n$ were observed at synchronization orders of $n =1, 2, 3$, and 4 (see Fig. 2C). Here, we should emphasize that a majority of the previous studies have reported noise-induced phase slips, where the slip interval has a statistical deviation, whereas such a perfectly periodic kinetic phase slip as shown here has been challenging to study in laboratory experiments (36-44). Especially for $n \geq 2$, referred to as fractional phase slip, it has been observed only in limited few devices to our knowledge (44 エラー! 参照元が見つかりません。). Both reasons are because they require strong nonlinearity with highly stable control of the synchronization parameters.

Our adiabatic amplitude modulation technique, moreover, allows us to verify the scaling law of the phase slip periods $\tau$ with respect to the depth of the Kuramoto potential $V_{\text{mod}}$. Because the synchronization originates in the saddle-node bifurcation, a $\tau^{-1} \propto \sqrt{1 - V_{\text{mod}}/V_C}$ dependence is predicted, with a critical depth of $V_C$ (43,45). Figure 2D shows the dependence on the synchronization order, $n =$1, 2, and 3, indicating good correspondence to the root dependence on $\tau^{-1}$. Such a bifurcation scaling for $n =$1 has been previously observed only in a few laser systems (36-39) because it requires the tunability of the Kuramoto potential to be highly stable. Here, we observed bifurcation scaling in the fractional phase slip ($n \geq 2$) for the first time in mesoscopic oscillators, owing to the well-engineered optomechanical nonlinearity. These results demonstrate the high stability and controllability of the optomechanical approach to manipulating the Kuramoto potential.

**Asymmetric Kuramoto potential via mixed-overtone modulation**

In addition to the synchronization with a single modulation tone, the Kuramoto potential can be modified to synthesize overtone potentials with two different orders. Here, we attempted to modulate the nonlinearity with two optical intensity modulation tones (see Fig. 3A). We used our control method to separately set the depth and phase for different overtone modulations and achieved a 1:2 mixed-overtone potential of $\zeta = \Delta\varphi + A_1 \cos(\varphi + \theta_1) + A_2 \cos 2\varphi$. This synthesized potential is asymmetric with two local minima having different stabilities. The fingerprint of the asymmetric Kuramoto potential was observed via the phase slip by setting the condition at the edge at the synchronization regime in a similar way as the previous experiments. When $A_1$ was zero, the $\pi$ phase slip had equal periods. On the other hand, double periods were observed when $A_1$ was finite (see Fig. 3B). This double period is a fingerprint of the asymmetric potential because the shallow (deep) potential causes the system to stay in synchronization a short (long) time.

To systematically probe the profile of the asymmetric Kuramoto potential, we demonstrated a bang-bang control protocol where the modulation was switched on and off in order to create and annihilate the Kuramoto potential alternately. When the modulation is switched off, the phase difference was uniformly distributed. Once the modulation was switched on, the phase difference was pulled into an equilibrium point ($\varphi_1$ or $\varphi_2$) at the local minimum in the potential (Fig. 3C). The equilibrium point in the deeper potential area ($\varphi_2$) pulled the phase difference with a higher probability than that in the shallow area ($\varphi_1$). Thus, the asymmetric potential profile could be verified by plotting the probability distribution of $\varphi_{\text{diff}}$ when the modulation is switched on. Note that the switching period is longer than any dynamics in the optomechanical system. Figure 3D shows the probability distribution of $\varphi_{\text{diff}}$ with respect to the modulation voltages for the first-order potential $V_1$. When $V_1 = 0$, a bimodal distribution appears because of the symmetry in the Kuramoto potential. On the other hand, as $V_1$ increases, the probability distribution becomes more asymmetric. In addition to the potential depth $A_1$, the relative phase offset $\theta_1$ also modulates the asymmetry. Figure 3E shows the probability distribution with respect to $\theta_1$. At $\theta_1 = 0$, the superposition of the first- and second-order potential does not break the symmetry; thus, a bimodal distribution appears. In contrast, the distribution is asymmetric for $\theta_1 \neq 0$ because of the symmetry breaking due to the superposition of the two potentials.

**Topology and non-reciprocity in synchronization path with dynamical potential synthesis**

A further extension of our idea is dynamical control of a synthesized Kuramoto potential. Here, the second- and the first-order potentials are synthesized with a dynamically modulated control phase $\theta_C(t) = 2\pi t/T$ as $\zeta(t) = \Delta\varphi_{\text{diff}} + A_1 \cos(\varphi_{\text{diff}} + \theta_C(t)) + A_2 \cos 2\varphi_{\text{diff}}$. This dynamical potential synthesis temporally changes the asymmetry of the Kuramoto potential and artificially induces a phase slip in which the potential stability transits from being stable to unstable (see Fig. 4A). The dynamical path of the synchronization is characterized by the total phase slip integrated along the closed loop $\gamma$ in the parameter space:

$$\Theta \equiv \oint_\gamma d\varphi_{\text{diff}}. \tag{3}$$

The parameter space can be mapped onto a toroidal surface where the major (minor) angle corresponds to $\theta_C(t)$ ($\varphi_{\text{diff}}$). Especially when the path is closed, the winding number, $W \equiv \Theta/2\pi$, can be defined to identify the topological properties of the dynamical path. Such topological properties in periodically modulated one-dimensional systems, i.e., 1+1 dimensional systems, have been discussed, an example being the Thouless pump in cold-atom and photonic systems (46-48).

When the two mechanical modes are completely phase synchronized without potential synthesis (i.e., $A_1 = 0$), we obtained a trivial path with $\Theta = 0$ ($W = 0$) (see Fig. 4B and 4C). On the other hand, the dynamical synthesis with a finite $A_1$ and temporal modulation $\theta_C(t)$ make a completely different topological path with respect to the ratio $A_1/A_2$. In the case where $|A_1/A_2| > 2$, a path characterized by $\Theta = 2\pi$ ($W = 1$) was obtained with composed of two slips, each begin $\pi$ phase slip (see Fig. 4D and 4E). Here, the initial and final phases are equivalent, due to the $2\pi$ phase slip in total, so that $W$ can be defined. Intuitively, this trajectory can be understood as that the large $A_1/A_2$ causes dragging of the phase along the first-order potential change. In contrast, when $|A_1/A_2| < 2$, the path has a quantized phase slip that occurs just once per period, i.e., $\Theta = \pi$ (see Fig. 4F and 4G). Because the final phase is not equivalent to the initial phase, the trajectory is not closed on the toroidal surface in which the winding number cannot be determined. This trajectory originates in the finite detuning, which causes the global tilt in the Kuramoto potential, and results in hysteresis in the path depending on the initial condition. Moreover, a nontrivial topological path is also observed in the case of the 1:3 mixed over-tone potential with the third-order potential given by $\zeta(t) = \Delta\varphi + A_1 \cos(\varphi + \theta_C(t)) + A_3 \cos 3\varphi$ (see Fig. 4H-4K). In addition to the path with $\Theta = 2\pi$ through the triple $2\pi/3$-quantized phase slips (see Fig. 4H and 4I), small detuning causes a quantized fractional phase slip of $\Theta = 4\pi/3$ due to the hysteresis (see Fig. 4J and 4K). These experimental results are in good agreement with the theoretical investigation of the Kuramoto model in the Supplementary Materials.

We found another distinct feature of these exotic topological paths with a fractional winding number due to hysteresis, i.e., a large non-reciprocity with respect to the sweep direction of $\theta_C(t)$. In the case of $\Theta = 2\pi$, the topological path does not show any non-reciprocity with respect to the sweep direction (see Fig. 5A and 5B). On the other hand, a large difference between the clockwise and counterclockwise paths was found in the case of $\Theta = \pi$. A different path appears at half of the control parameter range, i.e., $\Delta\theta_C \sim \pi$ (see Fig. 5C and 5D). Such a nonreciprocal feature was also found in the case of $\Theta = 4\pi/3$ with $\Delta\theta_C \sim 1.26\pi$ (see Fig 5E and 5F). The large value of $\Delta\theta_C$ features the non-reciprocal topological dynamics in the adiabatic control of synchronization.

**Discussion**

Synthesized synchronization via optomechanical Floquet engineering not only tailors the shape of the potential along the synchronization phase in the periodic one-dimensional space, but also adds one more periodic one-dimensional control parameter space, i.e., realizing a periodic 1+1-dimensional toroidal space. This situation is analogous to the Thouless pump in cold atomic systems and photonic systems consisting of one-dimensional array of atoms and optical modes in real space (46-48). It is noteworthy that our approach forms a periodic one-dimensional system in the phase space of two mechanical oscillators. In other words, by increasing the number of mechanical oscillators, the parameter space, i.e., the dimension of the topology, can be expanded with the use of $N+1$ oscillators to $N+1$ dimensions. Thus, our approach is universal, i.e., its applicability is not limited to multimode mechanical systems (49,50) but can be extended to systems embodying high-dimensional topological physics by incorporating tunable over-tone nonlinearity in multiple harmonic oscillators.

Since multistability in the Kuramoto potential can be precisely manipulated within the parameter space, it is expected that it can be exploited in an oscillator network that corresponds to high-dimensional information spaces such as clock operations. This would facilitate development of hardware for constructing a high-dimensional Kuramoto network that could be used in non-von Neumann computer architectures and multi-dimensional memories, applications in which mesoscopic solid-state oscillators have so far been difficult to implement. Furthermore, the ability to control individual components (orders) of the Kuramoto potential for multiple oscillators may help to uncover the origin of efficient information processing in complicated biological systems from the viewpoint of information thermodynamics and optimal control theory.

In conclusion, we have demonstrated tailor-made synchronization by synthesizing and adiabatically controlling the Kuramoto potential in an optomechanical resonator. We found that optical manipulation of the Kuramoto potential in mechanical oscillators could provide a highly controllable quantized phase slip at the edge of the synchronization regime and an asymmetric Kuramoto potential with mixed overtone modulations. Furthermore, we found that the adiabatic control of the Kuramoto potential revealed the topological properties of the dynamical paths in phase space and their giant non-reciprocity. We believe that synthesized Kuramoto potentials could be used as a reconfigurable and multi-functional building block in next-generation small-world networks that have advanced neuromorphic computation and information architectures or imitate complicated biological processes with well-engineered mesoscopic devices.

**Acknowledgments:**

This work was supported by JSPS KAKENHI (21H05020, 23H05463). The authors acknowledge constructive discussion with V. Bastidas.

1. J. Buck and E. Buck, Mechanism of Rhythmic Synchronous Flashing of Fireflies. *Science* 159, 1319 (1968).
2. L. Glass, Synchronization and rhythmic processes in physiology. *Nature* 410, 277 (2001).
3. K. Wadenpfuhl and C. S. Adams, Emergence of Synchronization in a Driven-Dissipative Hot Rydberg Vapor. *Phys. Rev. Lett.,* 131(14), 143002 (2023).

4. A. B. Cawthorne, P. Barbara, S. V. Shitov, C. J. Lobb, K. Wiesenfeld, and A. Zangwill, Synchronized oscillations in Josephson junction arrays: The role of distributed coupling, *Phys. Rev. B* 60, 7575 (1999).

5. K. M. Cuomo and A. V. Oppenheim, Circuit Implementation of Synchronized Chaos with Applications to Communications, *Phys. Rev. Lett.* 71, 65, (1993).

6. L. Kocarev and U. Parlitz, General Approach for Chaotic Synchronization with Applications to Communication, *Phys. Rev. Lett.* 74, 5028 (1995).

7. J. J. Hopfield and A.V. M. Herz, Rapid local synchronization of action potentials: Toward computation with coupled integrate-and-fire neurons. *Proc. Natl Acad. Sci. USA* 92, 6655–6662 (1995).

8. G. Filatrella, A. H. Nielsen, and N. F. Pedersen, Analysis of a power grid using a Kuramoto-like model. *Eur. Phys. J.* B 61, 485 (2008).

9. Y. Guo, D. Zhang, Z. Li, Q. Wang, and D. Yu, Overviews on the applications of the Kuramoto model in modern power system analysis. *International Journal of Electrical Power and Energy Systems* 129, 106804 (2021).

10. S.-B. Shim, M. Imboden, and P. Mohanty, Synchronized oscillation in coupled nanomechanical oscillators. *Science* 316, 95 (2007),

11. M. Zalalutdinov, K. Aubin, M. Pandey, A. Zehnder, R. Rand, H. Craighead, J. Parpia, and B. Houston, Frequency entrainment for micromechanical oscillator. *Appl. Phys. Lett.* 83, 3281 (2003).

12. M. H. Matheny, M. Grau, L. G. Villanueva, R. B. Karabalin, M. C. Cross, and M. L. Roukes, Phase Synchronization of Two Anharmonic Nanomechanical Oscillators. *Phys. Rev. Lett.* 112, 014101 (2014).

13. M. Zhang, G. S. Wiederhecker, S. Manipatruni, A. Barnard, P. McEuen, and M. Lipson, Synchronization of Micromechanical Oscillators Using Light. *Phys. Rev. Lett.* 109, 233906 (2012).

14. E. Gil-Santos, M. Labousse, C. Baker, A. Goetschy, W. Hease, C. Gomez, A. Lemaître, G. Leo, C. Ciuti, and I. Favero, Light-Mediated Cascaded Locking of Multiple Nano-Optomechanical Oscillators. *Phys. Rev. Lett.* 118, 063605 (2017).

15. C. C. Rodrigues, C. M. Kersul, A. G. Primo, M. Lipson, T. P. M. Alegre, and G. S. Wiederhecker, Optomechanical synchronization across multi-octave frequency spans. *Nat. Commun*. 12, 5625 (2021).

16. A. Hohl, A. Gavrielides, T. Erneux, and V. Kovanis, Localized Synchronization in Two Coupled Nonidentical Semiconductor Lasers, *Phys. Rev. Lett.* 78, 4745 (1997).

17. G. Kozyreff, A. G. Vladimirov, and P. Mandel, Global Coupling with Time Delay in an Array of Semiconductor Lasers, *Phys. Rev. Lett.* 85 3809 (2000).

18. P. Barbara, A. B. Cawthorne, S. V. Shitov, and C. J. Lobb, Stimulated Emission and Amplification in Josephson Junction Arrays, *Phys. Rev. Lett.* 82, 1963 (1999).

19. R. S. Shaikhaidarov, K. H. Kim, J. W. Dunstan, I. V. Antonov, S. Linzen, M. Ziegler, D. S. Golubev, V. N. Antonov, E. V. Il'ichev, and O. V. Astafiev, Quantized current steps due to the a.c. coherent quantum phase-slip effect, *Nature* 608, 45 (2022).

20. Y. Kuramoto, *International Symposium on Mathematical Problems in Theoretical Physics*, Lecture notes in Physics, 30, 420 (1975).

21. J. A. Acebrón, L. L. Bonilla, C. J. P. Vicente, F. Ritort, and R. Spigler, The Kuramoto model: A simple paradigm for synchronization phenomena. *Rev. Mod. Phys.* 77(1), 137-185 (2005).

22. Y. L. Maistrenko, B. Lysyansky, C. Hauptmann, O. Burylko, and P. A. Tass, Multistability in the Kuramoto model with synaptic plasticity. *Phys. Rev. E* 75, 066207 (2007).

23. T. Kotwal, X. Jiang, and D. M. Abrams, Connecting the Kuramoto Model and the Chimera State, *Phys. Rev. Lett.* 119, 264101 (2017).

24. M. Aspelmeyer, T. J. Kippenberg, and F. Marquardt, Cavity optomechanics, *Rev. Mod. Phys.* 86, 1391 (2014).

25. L. Mercadé, K. Pelka, R. Burgwal, A. Xuereb, A. Martínez, and E. Verhagen, Floquet phonon lasing in multimode optomechanical systems. *Phys. Rev. Lett.*, 127(7), 073601 (2021).

26. K. Pelka, G. Madiot, R. Braive, and A. Xuereb, Floquet control of optomechanical bistability in multimode systems. *Phys. Rev. Lett.* 129(12), 123603 (2022).

27. X. Dong, M. I. Dykman, and H. B. Chan, Strong negative nonlinear friction from induced two-phonon processes in vibrational systems. *Nat. Commun*. 9, 3241 (2018).

28. M. J. Seitner, M. Abdi, A. Ridolfo, M. J. Hartmann, and E. M. Weig, Parametric Oscillation, Frequency Mixing, and Injection Locking of Strongly Coupled Nanomechanical Resonator Modes, *Phys. Rev. Lett.* 118, 254301 (2017).

29. M. Sumetsky, Whispering-gallery-bottle microcavities: the three-dimensional etalon, *Opt. Lett.* 29, 1, 8 (2004).

30. M. Pöllinger, D. O'Shea, F. Warken, and A. Rauschenbeutel, Ultrahigh-$Q$ Tunable Whispering-Gallery-Mode Microresonator, *Phys. Rev. Lett.* 103, 053901 (2009).

31. M. Asano, Y. Takeuchi, W. Chen, Ş. K. Özdemir, R. Ikuta, N. Imoto, L. Yang, and T. Yamamoto, Observation of optomechanical coupling in a microbottle resonator. *Laser & Photon. Rev.* 10(4), 603-611 (2016).

32. A. J. R. MacDonald, B. D. Hauer, X. Rojas, P. H. Kim, G. G. Popowich, and J. P. Davis, Optomechanics and thermometry of cryogenic silica microresonators. *Phys. Rev. A* 93(1), 013836 (2016).

33. J. Sheng, X. Wei, C. Yang, and H. Wu, Self-organized synchronization of phonon lasers. *Phys. Rev. Lett.* 124(5), 053604 (2020).

34. J. S. Langer and V. Ambegaokar, Intrinsic resistive transition in narrow superconducting channels. *Physical Review*, 164(2), 498 (1967).

35. A. Bezryadin, C. N. Lau, and M. Tinkham, Quantum suppression of superconductivity in ultrathin nanowires. *Nature* 404(6781), 971-974 (2000).

36. K. V. Volodchenko, V. N. Ivanov, S-H Gong, M. Choi, Y-J Park, and C-M Kim, Phase synchronization in coupled Nd:YAG lasers, Opt. Lett. 26, 1406 (2001).

37. D. I. Kim, D-S Lee, Y-J Park, G. U. Kim, and C-M Kim, Phase synchronization of chaotic lasers, Opt. Express 14, 702 (2006).

38. L. Zhu, A. Raghu, and Y-C Lai, Experimental Observation of Superpersistent Chaotic Transients, *Phys. Rev. Lett.* 86, 4017 (2001).

39. S. Boccaletti, E. Allaria, R. Meucci, and F. T. Arecchi, Experimental Characterization of the Transition to Phase Synchronization of Chaotic CO2 Laser Systems, *Phys. Rev. Lett.* 89, 194101 (2002).

40. B. Kelleher, D. Goulding, S. P. Hegarty, G. Huyet, D. Y. Cong, A. Martinez, A. Lemaître, A. Ramdane, M. Fischer, F. Gerschütz, and J. Koeth, Excitable phase slips in an injection-locked single-mode quantum-dot laser. *Opt. Lett.* 34(4), 440-442 (2009).

41. M. Tortarolo, B. Lacoste, J. Hem, C. Dieudonné, M.-C. Cyrille, J. A. Katine, D. Mauri, A. Zeltser, L. D. Buda-Prejbeanu, and U. Ebels, Injection locking at 2f of spin torque oscillators under influence of thermal noise. *Sci Rep* 8(1), 1728 (2018).

42. R. Herrero, M. Figueras, F. Pi, and G. Orriols, Phase synchronization in bidirectionally coupled optothermal devices, *Phys. Rev. E* 66, 036223 (2002).

43. A. Pikovsky, G. Osipov, M. Rosenblum, M. Zaks, and J. Kurths, Attractor-repeller collision and eyelet intermittency at the transition to phase synchronization. *Phys. Rev. Lett.* 79(1), 47 (1997).

44. R. Lebrun, A. Jenkins, A. Dussaux, N. Locatelli, S. Tsunegi, E. Grimaldi, H. Kubota, P. Bortolotti, K. Yakushiji, J. Grollier, A. Fukushima, S. Yuasa, and V. Cros, Understanding of phase noise squeezing under fractional synchronization of a nonlinear spin transfer vortex oscillator. Phys. Rev. Lett. 115(1), 017201 (2015).

45. Z. Zheng, G. Hu, and B. Hu, Phase slips and phase synchronization of coupled oscillators. *Phys. Rev. Lett.* 81(24), 5318. (1998).

46. M. Lohse, C. Schweizer, O. Zilberberg, M. Aidelsburger, and I. Bloch, A Thouless quantum pump with ultracold bosonic atoms in an optical superlattice. *Nat. Phys.* 12(4), 350-354 (2016).

47. S. Nakajima, T. Tomita, S. Taie, T. Ichinose, H. Ozawa, L. Wang, M. Troyer, and Y. Takahashi, Topological Thouless pumping of ultracold fermions. *Nat. Phys.* 12(4), 296-300 (2016).

48. Y. E. Kraus, Y. Lahini, Z. Ringel, M. Verbin, and O. Zilberberg, Topological states and adiabatic pumping in quasicrystals. *Phys. Rev. Lett.* 109(10), 106402. (2012)

49. J. Doster, T. Shah, T. Fösel, P. Paulitschke, F. Marquardt, and E. M. Weig, Observing polarization patterns in the collective motion of nanomechanical arrays. *Nat. Commun.* 13, 2478 (2022).

50. J. J. Slim, C. C. Wanjura, M. Brunelli, J. del Pino, A. Nunnenkamp, E. Verhagen, Optomechanical realization of the bosonic Kitaev chain. *Nature* 627, 767 (2024).

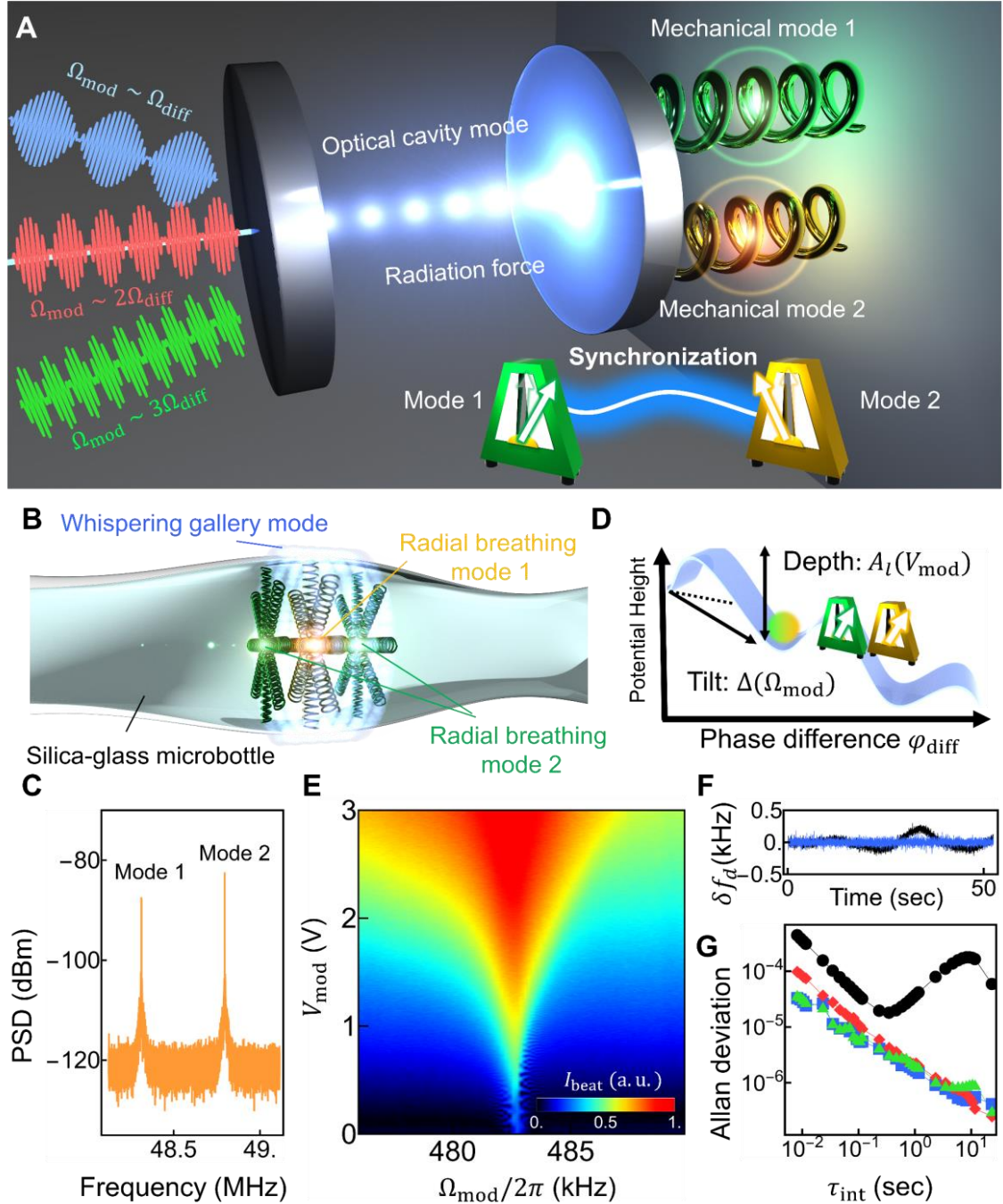

**Fig. 1. Formation of Kuramoto potential using a cavity optomechanical system.** (**A**) Conceptual illustration of synchronization between two mechanical oscillators via cavity optomechanical Floquet engineering. (**B**) Schematic diagram of silica-glass microbottle resonator. (**C**) Two-mode mechanical oscillation spectra in the microbottle resonator. (**D**) Illustration of Kuramoto potential controlled by parameters: potential depth, $A_l$, and potential tilt, $\Delta$. (**E**) Color map of normalized beat-note amplitude, $I_{beat}$, between two mechanical modes. The strong beat note amplitude (red colored area) corresponds to phase synchronization. (**F**) Temporal fluctuation of frequency difference $\delta f_d$ without synchronization (black) and with synchronization (blue). (**G**) Frequency stability evaluated by Allan deviation with respect to the integration time, $\tau_{int}$. The black circles show the Allan deviation in the free-running two-mode oscillation, and the blue, red, green plots correspond to the Allan deviation in first-, second-, and third-order synchronization.

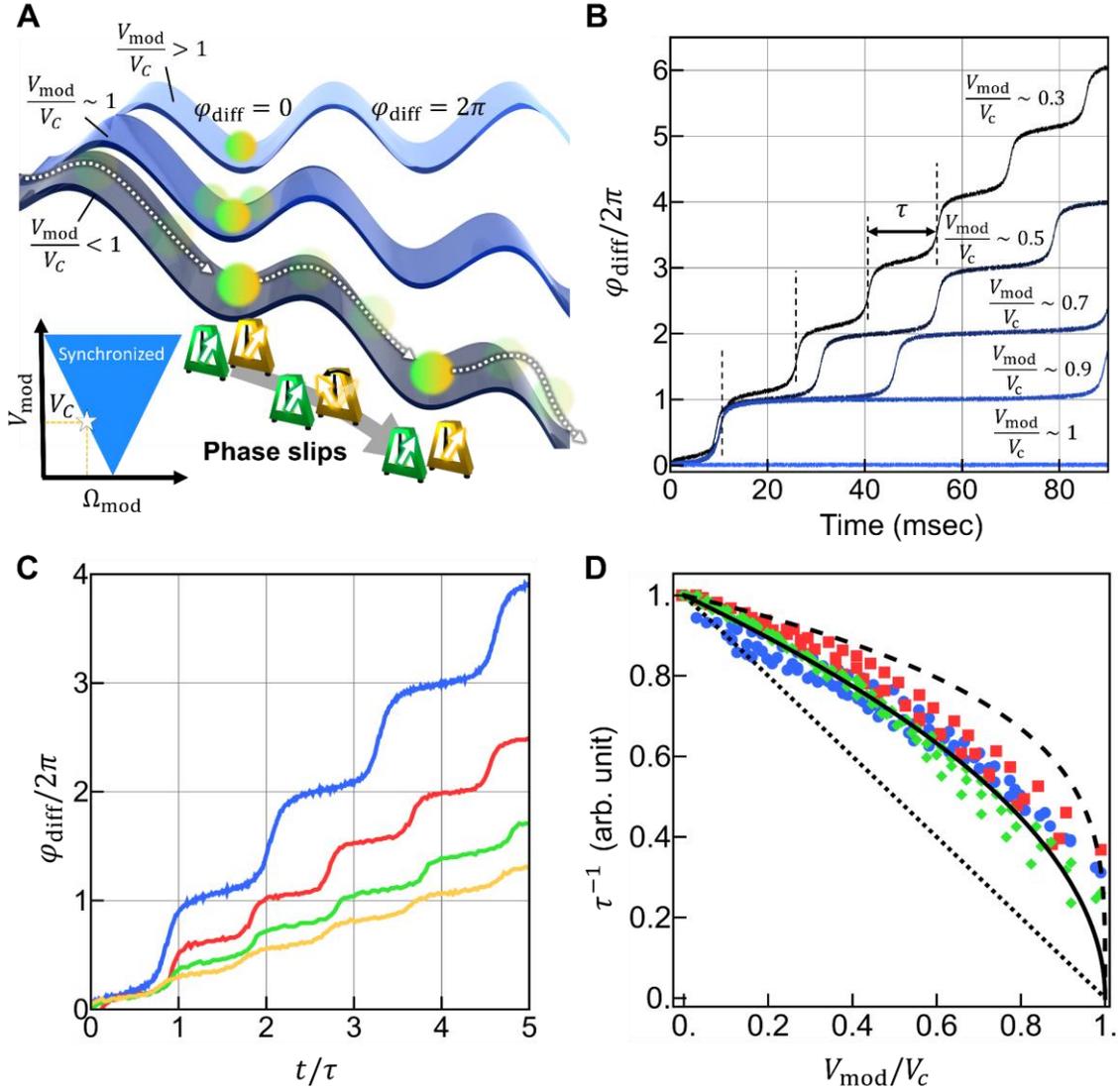

**Fig. 2. Quantized phase slip in optomechanical synchronization** (**A**) Conceptual illustration of quantized phase slip at the edge of the synchronization regime. (**B**) Quantized phase slip with $n = 1$ at different modulation voltages. (**C**) Quantized phase slip with $n = 1$ (blue), $n = 2$ (red), $n = 3$ (green), and $n = 4$ (yellow). (**D**) Inverse phase slip duration $\tau^{-1}$ with respect to the normalized synchronization strength $V_{\text{mod}}/V_c$. The blue circles, red squares, green diamonds correspond to the different-order synchronizations $n = 1, 2,$ and $3$. The solid, dashed, dotted black lines are the theoretical curves of $\left(1 - \frac{V_{mod}}{V_c}\right)^\alpha$ with $\alpha = 1/2, 1/4,$ and $1$.

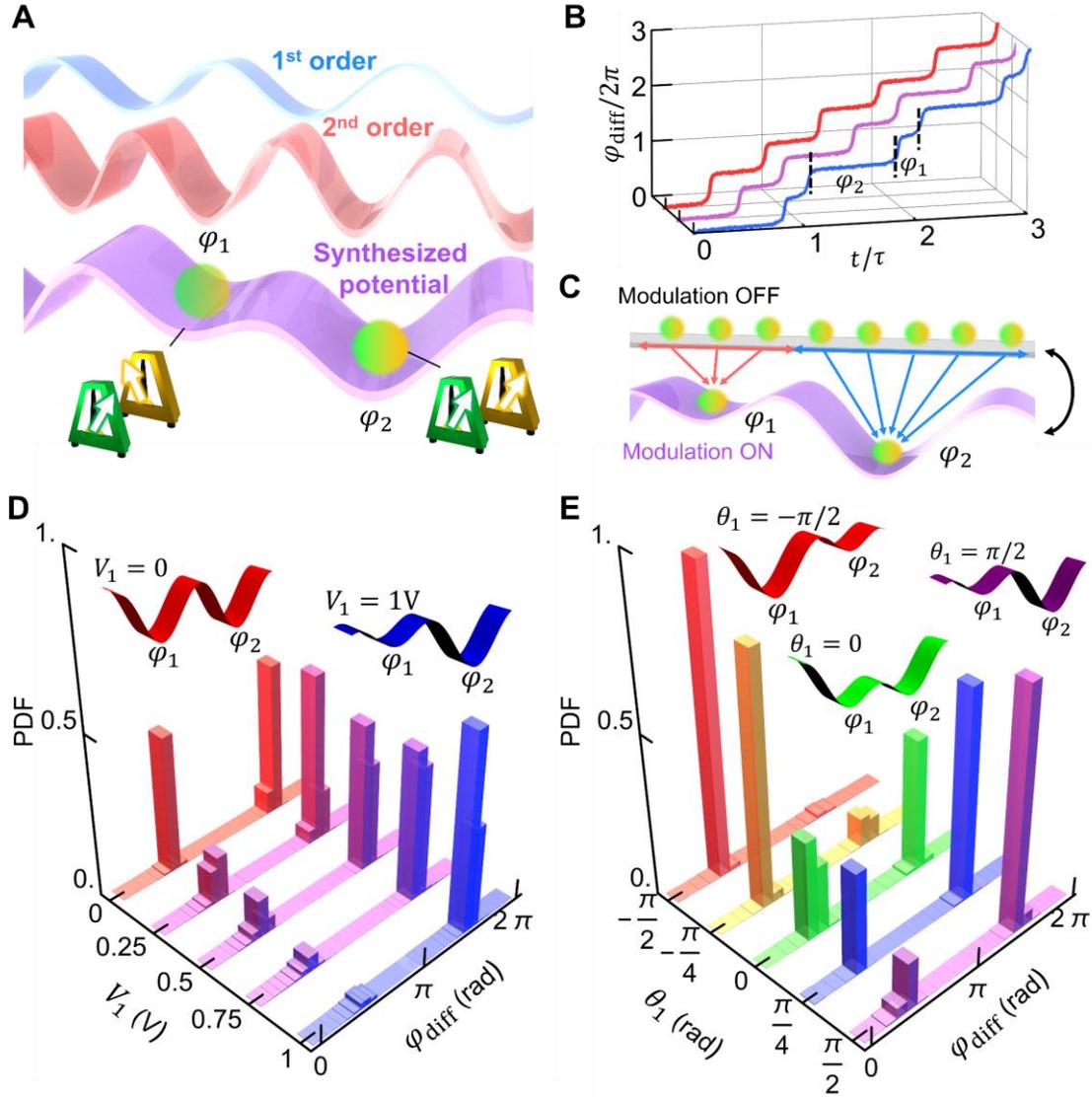

**Fig. 3. Asymmetric Kuramoto potentials** (**A**) Illustration of static synthesis of Kuramoto potential with first- and second-order Kuramoto potentials colored blue and red, respectively. The synthesized potential (purple) has an asymmetric profile and shows bistability at $\varphi_1$ and $\varphi_2$. (**B**) $\pi$-quantized phase slip with asymmetric phase potential where $\theta_1 = \pi/2$ and $V_1 = 0$ V (red), 0.25 V (purple), and 1 V (blue). A longer (shorter) plateau corresponds to a deeper (shallower) potential. (**C**) Illustration of bang-bang control protocol to switch on and off the modulation to create and annihilate the asymmetric potential. (**D**) Probability density function (PDF) of $\varphi_{\text{diff}}$ with respect to $V_1$ with $\theta_1 = \pi/2$. (**E**) PDF of $\varphi_{\text{diff}}$ with respect to $\theta_1$ with $V_1 = 1$ V. The insets in (**D**) and (**E**) show the potential profiles.

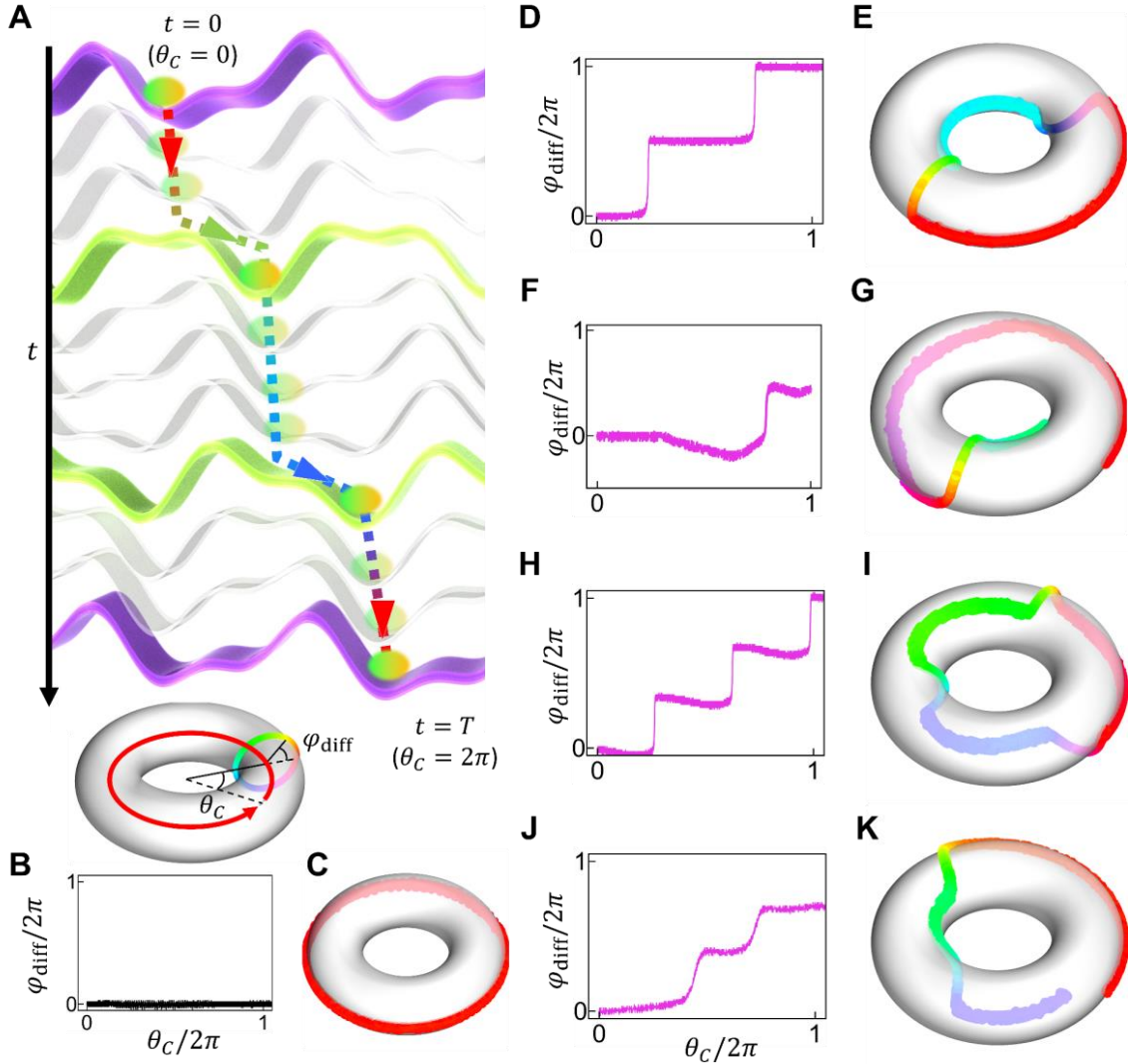

**Fig. 4. Topological synchronization path in dynamically synthesized Kuramoto potential** (**A**) Illustration of dynamical synthesis of the Kuramoto potential. The trajectory $\varphi_{\text{diff}}$ can be mapped to a toroidal parameter space with the control phase $\theta_C$ (**B**) and (**C**) Trivial path in the toroidal parameter space in the case of synchronization without dynamical synthesis. (**D**) and (**E**) Topological path with $\Theta = 2\pi$ in the 1:2 mixed-overtone potential. (**F**) and (**G**) Topological path with $\Theta = \pi$ in the 1:2 mixed-overtone potential. (**H**) and (**I**) Topological path with $\Theta = 2\pi$ in the 1:3 mixed-overtone potential. (**J**) and (**K**) Topological path with $\Theta = 4\pi/3$ in the 1:3 mixed-overtone potential.

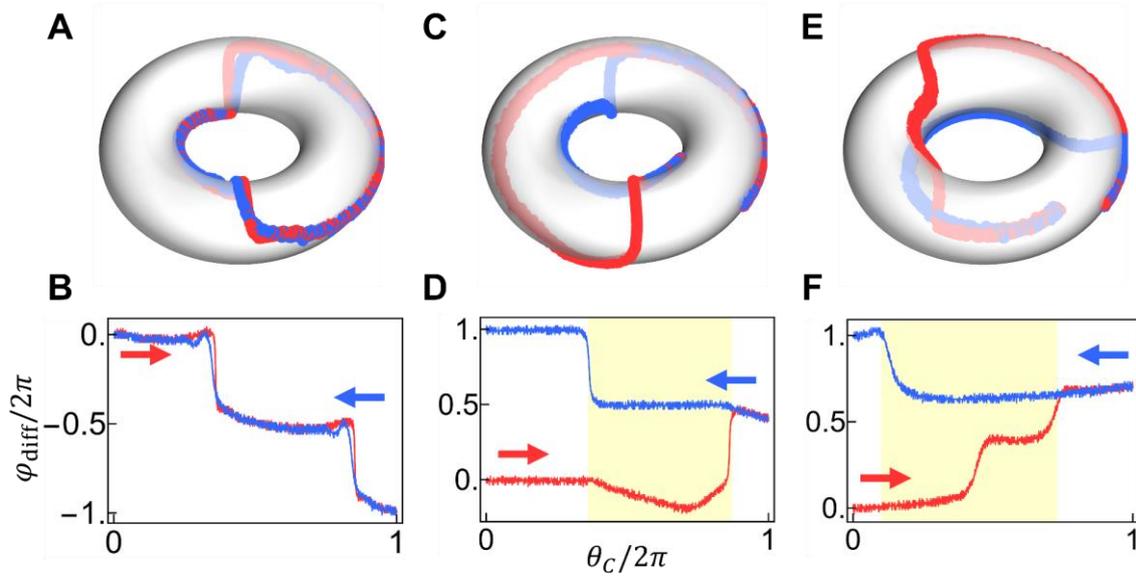

**Fig. 5. Non-reciprocity in the topological path.** The topological path with sweeping $\theta_C$ in the clockwise (red) and counterclockwise (blue) directions in a 1:2 mixed-overtone potential with $\Theta = 2\pi$ [(**A**) and (**B**)] and $\Theta = \pi$ [(**C**) and (**D**)], and a 1:3 mixed-overtone potential with $\Theta = 4\pi/3$ [(**E**) and (**F**)]. The yellow shaded area corresponds to the non-reciprocal path where the duration $\Delta\theta_C \sim \pi$ and $1.26\pi$ for the 1:2 and 1:3 mixed-overtone potential, respectively.

# Supplementary Materials

**Theory of two-mode mechanical phase synchronization via optomechanical Floquet engineering**

Let us theoretically describe two-mode mechanical phase synchronization via the optical intensity modulation approach. The equations of motion of the optical amplitude in a cavity, $a$, and the mechanical displacement in the $j$th mode, $x_j$ ($j = 1,2$), are given by

$$\dot{a} = \left[i\Delta_{\text{opt}} - \frac{\kappa}{2} + i(g_1 x_1 + g_2 x_2)\right] a + \sqrt{\kappa_{\text{in}}} a_{\text{in}}, \tag{S1}$$

$$\ddot{x}_1 + \Gamma_1 \dot{x}_1 + \Omega_1^2 x_1 = -\frac{\hbar g_1}{m_{\text{eff}}} |a|^2, \tag{S2}$$

$$\ddot{x}_2 + \Gamma_2 \dot{x}_2 + \Omega_2^2 x_2 = -\frac{\hbar g_2}{m_{\text{eff}}} |a|^2, \tag{S3}$$

where $\Delta_{\text{opt}}, \kappa, \kappa_{\text{in}}, \Gamma_j$, and $\Omega_j$ are the optical cavity detuning, optical dissipation rate, cavity input rate, mechanical dissipation rates, and mechanical eigen frequencies, respectively, of the $j$th mechanical mode ($j = 1,2$). The laser input amplitude is denoted by $a_{\text{in}}$. For simplicity, we will assume that the two mechanical resonators have the almost same effective mass $m_{\text{eff},1} \sim m_{\text{eff},2} \sim m_{\text{eff}}$. The optical mode and mechanical modes are mutually coupled via dispersive optomechanical couplings whose coupling strengths are given as $g_j = x_{\text{zpf},j} \partial \omega_{\text{opt}} / \partial x_j$, where $x_{\text{zpf},j}$ is the zero-point fluctuation, $\partial \omega_{\text{opt}} / \partial x_j$ is the optomechanical responsivity of the $j$th mode, and $\omega_{\text{opt}}$ is the eigen angular frequency of the optical cavity. Note that, to simply provide insight for the nonlinear Floquet engineering for mechanical modes, we will restrict the discussion to optical cavity modes in a steady state ($\dot{a} = 0$) in which the two mechanical modes have a finite delay $\tau_d$ (i.e., $x_j(t - \tau_d)$) (S1). Thus, we obtain

$$\begin{aligned}
-\frac{\hbar}{m_{\text{eff}}} |a|^2 &= -\frac{\hbar}{m_{\text{eff}}} \frac{\kappa_{\text{in}}}{(\Delta_{\text{opt}} + g_1 x_1 + g_2 x_2)^2 + \frac{\kappa^2}{4}} |a_{\text{in}}(t)|^2, \\
&\sim -f(t) \sum_{m=0}^{\infty} \eta_m \left(g_1 x_1(t - \tau_d) + g_2 x_2(t - \tau_d)\right)^m, \\
&= -f(t) \sum_{m=0}^{\infty} \sum_{k=0}^{m} \frac{m!}{k!(m-k)!} \eta_m g_1^k g_2^{m-k} x_1^k(t - \tau_d) x_2^{m-k}(t - \tau_d), \\
&\sim -f(t) \sum_{m=0}^{\infty} \sum_{k=0}^{m} \tilde{\eta}_{m,k} x_1^k(t - \tau_d) x_2^{m-k}(t - \tau_d),
\end{aligned} \tag{S5}$$

where $f(t) = 1 + \sin(\epsilon \cos \Omega_{\text{mod}} t)$ is the temporal modulation of forces optically exerted on the mechanical modes with modulation depth $\epsilon$ and frequency $\Omega_{\text{mod}}$, $\eta_m$ is the coefficient in the Taylor expansion, and $\tilde{\eta}_{m,k} = \frac{m!}{(k!(m-k)!)} g_1^k g_2^{m-k} \eta_m$ is the total nonlinear coefficient. Moreover, because $\epsilon \ll 1$, the dynamical modulation part can be approximated as $f(t) \sim (1 + \epsilon \sin \Omega_{\text{mod}} t)$. Here, we assume that the optical mode is in the steady state with a finite delay $\tau_d$, whose analytical expression is discussed later. The cross term of the displacement can be approximated to

$$\begin{aligned}
x_1^k(t - \tau_d) x_2^{m-k}(t - \tau_d) &\sim [x_1^k(t) + k\tau_d x_1^{k-1} \dot{x}_1(t)][x_2^{m-k}(t) + (m-k)\tau_d x_2^{m-k-1} \dot{x}_2(t)], \\
&\sim x_1^k x_2^{m-k} + \tau_d\left(k x_1^{k-1} \dot{x}_1 x_2^{m-k} + (m-k) x_2^{m-k-1} \dot{x}_2 x_1^k\right),
\end{aligned} \tag{S6}$$

where the first and second approximations are made by assuming $|\tau_d \dot{x}_j| \ll |x_j|$ ($j = 1,2$). Because the non-conservative terms, $k\tau_d(x_1^{k-1}\dot{x}_1 x_2^{m-k} + x_2^{m-k-1}\dot{x}_2 x_1^k)$, include a delay factor $\tau_d$, we take into account the two essential dissipative terms in the nonlinear dynamics: optomechanical gain force $F_{G,j}$ and nonlinear dissipation force $F_{\text{NLD},j}$ given by

$$F_{G,j} \sim -g_j^2 \tau_d \eta_1 \dot{x}_j, \tag{S7}$$

$$\begin{aligned} F_{\text{NLD},1} &\sim -g_1 \tau_d f(t)\big(3\tilde{\eta}_{3,3} x_1^2 + \tilde{\eta}_{3,1} x_2^2\big)\dot{x}_1 \sim -\gamma_1 x_1^2 \dot{x}_1, \\ F_{\text{NLD},2} &\sim -g_2 \tau_d f(t)\big(3\tilde{\eta}_{3,0} x_2^2 + \tilde{\eta}_{3,2} x_1^2\big)\dot{x}_2 \sim -\gamma_2 x_1^2 \dot{x}_2, \end{aligned} \tag{S8}$$

where $\gamma_j \equiv 18\eta_3 g_j \tau_d$ is the nonlinear dissipation constant and we renormalize the linear mechanical dissipation rate by $\Gamma'_j = \Gamma_j - g_j^2 \tau_d \eta_1$, where the second term corresponds to the optomechanical gain. Thus, to discuss the self-oscillation regime in our experiment, we have to impose on the time delay that $\tau_d > \max_{j=1,2} \Gamma_j/(g_j \eta_1)$.

By transforming $x_j(t) = b_j(t)^{-i\Omega_j t} + c.c.$ where $c.c.$ denotes the complex conjugates, the slowly varying approximation to Eq. (S5) in the complex amplitude, $\dot{b}_j \ll \Omega_j b_j$, gives the equation of motion of complex amplitude as follows:

$$\dot{b}_j = -\frac{\Gamma'_j}{2} b_j + \gamma_j |b_j|^2 b_j - i\frac{g_j f(t) e^{i\Omega_j t}}{2\Omega_j} \sum_{m=0}^{\infty} \sum_{k=0}^{m} \tilde{\eta}_{m,k} x_1^k x_2^{m-k}, \tag{S9}$$

where the first and second terms on the right-hand side correspond to the renormalized linear and nonlinear dissipation in Eq. (S8). Here, we consider a generalized intensity modulation with modulation frequency, $\Omega_{\text{mod}} \sim r(\Omega_1 - \Omega_2) + \delta$, where $r$ is an integer and $\delta$ denotes the tunable detuning. The summation in the third term on the right-hand side, i.e., optomechanical conservative force, can be formally described as

$$F_j \equiv \sum_{m=0}^{\infty} \sum_{k=0}^{m} \tilde{\eta}_{m,k} x_1^k x_2^{m-k} = \sum_l \sum_n f_{l,n}(\mathbf{b}) e^{i(l\Omega_1 + n\Omega_2)t}. \tag{S10}$$

Thus, the total term can be expanded to

$$\begin{aligned} f(t) F_j e^{i\Omega_1 t} &= \bigg(1 + \frac{\epsilon}{2i}\big[e^{ir(\Omega_1-\Omega_2)t+i\delta t} - e^{-ir(\Omega_1-\Omega_2)t-i\delta t}\big]\bigg)\sum_l \sum_n f_{l,n}(\mathbf{b}) e^{i((l+1)\Omega_1 + n\Omega_2)t} \\ &\sim f_{-1,0} - \frac{\epsilon}{2i} f_{r-1,-r}(\mathbf{b}) e^{-i\delta t}, \end{aligned} \tag{S11}$$

$$\begin{aligned} f(t) F_j e^{i\Omega_2 t} &= \bigg(1 + \frac{\epsilon}{2i}\big[e^{ir(\Omega_1-\Omega_2)t+i\delta t} - e^{-ir(\Omega_1-\Omega_2)t-i\delta t}\big]\bigg)\sum_l \sum_n f_{l,n}(\mathbf{b}) e^{i(l\Omega_1 + (n+1)\Omega_2)t} \\ &\sim f_{0,-1} + \frac{\epsilon}{2i} f_{-r,r-1}(\mathbf{b}) e^{i\delta t}. \end{aligned} \tag{S12}$$

Note that each approximation extracts the lowest order of $l$ and $n$ in $f_{l,n}$. The static part of the optomechanical force can be written as $f_{-1,0} = \tilde{\eta}_{1,1} b_1 + \tilde{\eta}_{3,3} |b_1|^2 b_1$ and $f_{0,-1} = \tilde{\eta}_{0,1} b_2 + \tilde{\eta}_{3,3} |b_2|^2 b_2$ by taking into account of the third-order nonlinearity as well as the dissipation term. The equation of motion reduces to

$$\dot{b}_1 \sim -i\delta_1 b_1 - \frac{\Gamma'_1}{2} b_1 + \gamma_1 |b_1|^2 b_1 - i\alpha_1 |b_1|^2 b_1 + \frac{g_1 \epsilon}{4\Omega_1} f_{r-1,-r} e^{-i\delta t}, \tag{S13}$$

$$\dot{b}_2 \sim -i\delta_2 b_2 - \frac{\Gamma'_2}{2} b_2 + \gamma_2 |b_2|^2 b_2 - i\alpha_2 |b_2|^2 b_2 - \frac{g_2 \epsilon}{4\Omega_2} f_{-r,r-1} e^{i\delta t}. \tag{S14}$$

From these equations, the equation of motion in the mechanical amplitude and phase can be found by decomposing (S13) and (S14) by setting $b_j = B_j e^{-i\varphi_j}$ and taking the real and imaginary part of the equation as follows:

$$\dot{B}_1 \sim -\frac{\Gamma'_1}{2} B_1 + \gamma_1 B_1^2 B_1 + \frac{g_1 \epsilon}{4\Omega_1} \tilde{f}_{r-1,-r}(\boldsymbol{B}) \cos[r(\varphi_1 - \varphi_2) + \delta t], \tag{S15}$$

$$\dot{\varphi}_1 \sim \delta_1 + \alpha_1 B_1^2 + \frac{g_1 \epsilon}{4\Omega_1 B_1} \tilde{f}_{r-1,-r}(\boldsymbol{B}) \sin[r(\varphi_1 - \varphi_2) + \delta t], \tag{S16}$$

$$\dot{B}_2 \sim -\frac{\Gamma'_2}{2} B_2 + \gamma_2 B_2^2 B_2 - \frac{g_2 \epsilon}{4\Omega_2} \tilde{f}_{-r,r-1}(\boldsymbol{B}) \cos[r(\varphi_1 - \varphi_2) + \delta t], \tag{S17}$$

$$\dot{\varphi}_2 \sim \delta_2 + \alpha_2 B_2^2 + \frac{g_2 \epsilon}{4\Omega_2 B_2} \tilde{f}_{-r,r-1}(\boldsymbol{B}) \sin[r(\varphi_1 - \varphi_2) + \delta t], \tag{S18}$$

where $f_{l,m}(\boldsymbol{b}) = \tilde{f}_{l,m}(\boldsymbol{B}) e^{-il\varphi_1 - im\varphi_2}$ in (S12) and (S13). Assuming the steady states in the amplitude equation via self-oscillation (i.e., $\dot{B}_j = 0$ for $j = 1,2$), we obtain

$$B_1^2 \sim \frac{\Gamma'_1}{2\gamma_1} - \frac{g_1 \epsilon}{4\Omega_1 \gamma_1 B_1} \tilde{f}_{r-1,-r}(\boldsymbol{B}) \cos[r(\varphi_1 - \varphi_2) + \delta t], \tag{S19}$$

$$B_2^2 \sim \frac{\Gamma'_2}{2\gamma_2} + \frac{g_2 \epsilon}{4\Omega_2 \gamma_2 B_2} \tilde{f}_{-r,r-1}(\boldsymbol{B}) \cos[r(\varphi_1 - \varphi_2) + \delta t]. \tag{S20}$$

Substituting them into the equation of motion of the phase, we obtain

$$\begin{aligned}\dot{\varphi}_1 \sim \tilde{\delta}_1 &- \frac{\alpha_1 g_1 \epsilon}{4\Omega_1 \gamma_1 B_1} \tilde{f}_{r-1,-r}(\boldsymbol{B}) \cos[r(\varphi_1 - \varphi_2) + \delta t] \\ &+ \frac{g_1 \epsilon}{4\Omega_1} \tilde{f}_{r-1,-r}(\boldsymbol{B}) \sin[r(\varphi_1 - \varphi_2) + \delta t],\end{aligned} \tag{S21}$$

$$\begin{aligned}\dot{\varphi}_2 \sim \tilde{\delta}_2 &+ \frac{\alpha_2 g_2 \epsilon}{4\Omega_2 \gamma_2 B_2} \tilde{f}_{-r,r-1}(\boldsymbol{B}) \cos[r(\varphi_1 - \varphi_2) + \delta t] \\ &+ \frac{g_2 \epsilon}{4\Omega_2} \tilde{f}_{-r,r-1}(\boldsymbol{B}) \sin[r(\varphi_1 - \varphi_2) + \delta t].\end{aligned} \tag{S22}$$

Here, we should note that the third terms in Eqs (21) and (22) vanish by taking account the phase reduction $B_1 \sim B_2 \sim$ const. Finally, we obtain the Adler equation,

$$\dot{\varphi}_{\text{diff}} = \dot{\theta}_1 - \dot{\theta}_2 \sim \widetilde{\Delta} + K_0 \cos[r\varphi_{\text{diff}}], \tag{S23}$$

with

$$\widetilde{\Delta} = -\frac{\delta}{r} + \tilde{\delta}_1 - \tilde{\delta}_2, \tag{S24}$$

$$K = -\left( \frac{\alpha_1 g_1 \epsilon}{4\Omega_1 \gamma_1 B_1} \tilde{f}_{r-1,-r}(\boldsymbol{B}) + \frac{\alpha_2 g_2 \epsilon}{4\Omega_2 \gamma_2 B_2} \tilde{f}_{-r,r-1}(\boldsymbol{B}) \right). \tag{S25}$$

**Phase dynamics in a generalized Kuramoto potential**

Here, we provide a universal description for the time evolution of the difference in phase between two modes under a generalization of a standard Kuramoto potential in (S2 and S3). The generalized phase equation is given by modifying Eq. (1) in the main text to

$$\dot{\varphi}_{\text{diff}} = -\frac{\partial \zeta}{\partial \varphi_{\text{diff}}}, \qquad \zeta(\varphi_{\text{diff}}) = \Delta \varphi_{\text{diff}} + f(\varphi_{\text{diff}}), \tag{S26}$$

or by introducing a force field $g(\varphi_{\text{diff}})$,

$$\frac{d\varphi_{\text{diff}}}{dt} = g(\varphi_{\text{diff}}), \qquad g(\varphi_{\text{diff}}) = -\frac{\partial \zeta}{\partial \varphi_{\text{diff}}} = -\Delta - f'(\varphi_{\text{diff}}). \tag{S27}$$

We can easily integrate Eq. (S27) as

$$\int_{\varphi_i}^{\varphi_f} \frac{d\varphi_{\text{diff}}}{g(\varphi_{\text{diff}})} = t_f - t_i. \tag{S28}$$

Here, $\varphi_f$ and $\varphi_i$ are the phase difference at time $t_f$ and $t_i$, respectively. This equation gives the time required for $\varphi_{\text{diff}}$ to change from $\varphi_i$ to $\varphi_f$ under the phase interaction governed by the Kuramoto potential, $\zeta(\varphi_{\text{diff}})$. The left-hand side (l.h.s.) is finite if $g(\varphi_{\text{diff}})$ does not vanish in the integration region, but it logarithmically diverges at the point where $g(\varphi_{\text{diff}})$ vanishes.

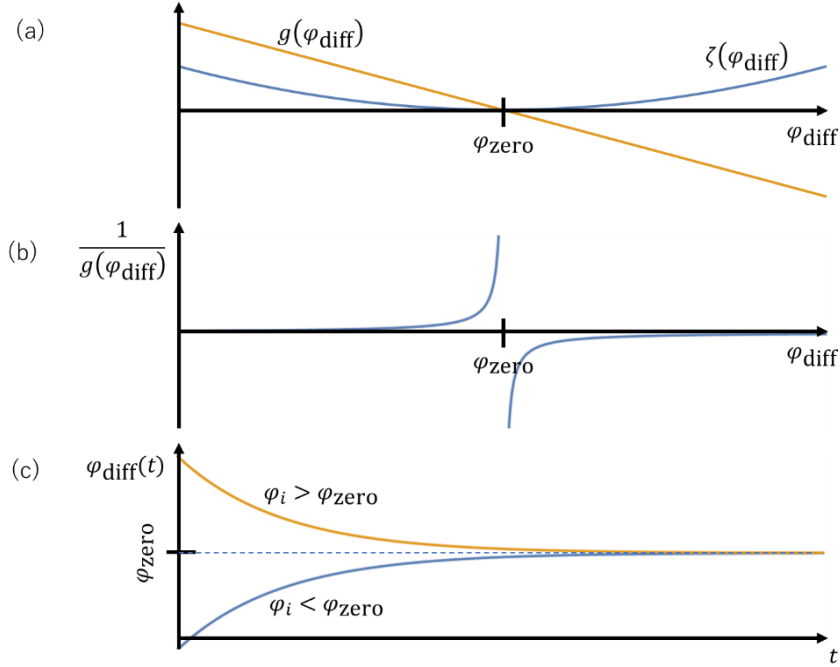

Fig. S1 (a) Example of a Kuramoto potential $\zeta(\varphi_{\text{diff}})$ which has a local minimum at $\varphi_{\text{zero}}$, and derived force field $g(\varphi_{\text{diff}})$, linearly crossing zero at $\varphi_{\text{zero}}$. (b) The integrand $g(\varphi_{\text{diff}})^{-1}$, which diverges at $\varphi_{\text{zero}}$. (c) The time evolution of the phase difference $\varphi_{\text{diff}}$ obtained by integrating equation (S28). The phase difference is stabilized at $\varphi_{\text{zero}}$ under both initial conditions, $\varphi_i < \varphi_{\text{zero}}$ and $\varphi_i > \varphi_{\text{zero}}$, inducing phase synchronization.

For example, if $\zeta(\varphi_{\text{diff}})$ takes the quadratic form $\zeta(\varphi_{\text{diff}}) = \frac{A}{2}(\varphi_{\text{diff}} - \varphi_{\text{zero}})^2$ with $A > 0$, then $g(\varphi_{\text{diff}})$ is approximately linear at the point it crosses the axis, i.e., $\varphi_{\text{zero}}$ (see Fig. S1 (a))

$$g(\varphi_{\text{diff}}) \sim -A(\varphi_{\text{diff}} - \varphi_{\text{zero}}). \tag{S29}$$

For $\varphi_i$ slightly lower than $\varphi_{\text{zero}}$, the integration over the negative neighborhood of $\varphi_{\text{zero}}$ is given by

$$\int_{\varphi_i}^{\varphi_{\text{zero}}-\varepsilon} \frac{d\varphi_{\text{diff}}}{g(\varphi_{\text{diff}})} \sim -A^{-1} \int_{\varphi_i-\varphi_{\text{zero}}}^{-\varepsilon} \frac{d\hat{\varphi}_{\text{diff}}}{\hat{\varphi}_{\text{diff}}} \xrightarrow[\varepsilon \to +0]{} A^{-1} \log \varepsilon. \tag{S30}$$

Here, we replace the integration variable by $\hat{\varphi}_{\text{diff}} = \varphi_{\text{diff}} - \varphi_{\text{zero}}$. This result indicates that *an infinitely long time is required for $\varphi_{\text{diff}}$ to approach $\varphi_{\text{zero}}$* (see Fig. S1 (c)). In other words, the root of $g(\varphi_{\text{diff}}) = 0$ (i.e. the local minima of the Kuramoto potential $\zeta(\varphi_{\text{diff}})$, see Fig. S1 (a)) is a stable point and $\varphi_{\text{diff}}$ approaches and is finally locked at the point. Because $g(\varphi_{\text{diff}})$ is the derivative of $\zeta(\varphi_{\text{diff}})$, the results indicate that *the phase difference is locked at* *the local minimum of the Kuramoto potential* $\zeta(\varphi_{\text{diff}})$. This is the basic mechanism to induce phase synchronization between two oscillators. It can be easily shown that $\varphi_{\text{diff}}$ departs from $\varphi_{\text{zero}}$ when $A < 0$. This means that the local maximum of the Kuramoto potential is unstable. Phase stability is guaranteed only at the local minimum.

Here, it should be noted that there is a difference from the dynamics induced by a standard equation of motion for a point particle under a conservative force. In that case, the integral of the equation of motion is given by

$$\sqrt{\frac{m}{2}} \int_{q_i}^{q_f} \frac{dq}{\sqrt{E - V(q)}} = t_f - t_i. \tag{S31}$$

Here $q$ is the position of a point particle, $E$ and $m$ are the total energy and mass of the particle, respectively, and $V(q)$ is the potential energy. Because of the square root in the denominator, the l.h.s. does not become infinite even for the zero kinetic-energy (i.e. zero velocity) position. This induces the motion of the particle to be reflected at the zero kinetic-energy position within a finite time duration, and the zero-denominator position is unstable. The particle at the potential minima is not stabilized by the kinetics but dissipation is required to push the system into the potential minimum by reducing the total energy, $E$. In contrast, the stability in our generalized Kuramoto model is simply governed by the system kinetics and dissipation is not required to push the system into the potential minimum.

**Examples of the Kuramoto potential**

Here, we show some examples of phase dynamics under simple potential shapes. The first example is the standard Kuramoto model. The potential is given by $\zeta = \Delta \varphi_{\text{diff}} + A \cos \varphi_{\text{diff}}$. The function $\zeta$ has the shape of a tilted washboard potential (top graph in Fig. S2) and has local minima when the potential tilt $\Delta$ is smaller than the depth $|A|$, i.e. $\Delta \leq |A|$ (green curve in Fig. S2). This condition gives the onset of synchronization, and the Arnold tongue is the triangular area satisfying this condition. As shown in Fig. S2, the integrand $1/g(\varphi_{\text{diff}})$ diverges at the local minima and an infinitely long time is required to arrive at the local minima. The local minima in $\zeta(\varphi_{\text{diff}})$ maintains the stability of synchronization, but the further tilt induced by the frequency detuning causes instability. When $\Delta > |A|$, a local minimum does not exist, and the synchronization is not maintained (orange and blue curves in Fig.S2). Under this condition, a finite time is enough for $\varphi_{\text{diff}}$ to pass over the local minima of $g(\varphi_{\text{diff}})$ and the periodic phase slip emerges.

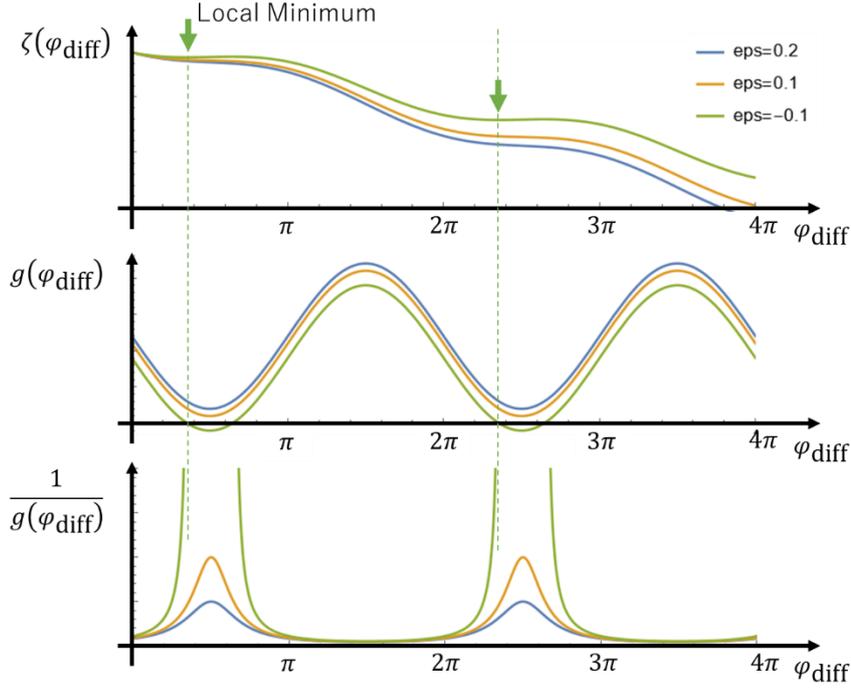

Fig. S2 $\zeta(\varphi_{\text{diff}})$, $g(\varphi_{\text{diff}})$, and the integrand $1/g(\varphi_{\text{diff}})$, for a standard sinusoidal Kuramoto potential with a detuning. "eps" is the detuning parameter, $\varepsilon$, defined by $\Delta = |A|(1 + \varepsilon)$. $\varepsilon < 0$ (green curve) corresponds to the condition $\Delta < |A|$, where the potential has a local minimum (shown by green arrows and dashed lines), the force field $g(\varphi_{\text{diff}})$ has zero values, and the integrand diverges, inducing phase synchronization. $\varepsilon > 0$ corresponds to the condition $\Delta > |A|$ (orange and blue curves), where no local minimum exists in $\zeta(\varphi_{\text{diff}})$ and a periodic phase slip is induced.

Next, we show this phase slip dynamics in the outer neighborhood of the synchronization onset by performing a numerical calculation. When the detuning $|\Delta|$ is slightly larger than $|A|$, i.e. $\Delta = A(1 + \varepsilon)$ with $0 < \varepsilon \ll 1$ (orange and blue curves in Fig.S2), the integration (S28) is given by

$$t_f - t_i = -A^{-1} \int_{\varphi_i}^{\varphi_f} \frac{d\varphi_{\text{diff}}}{1 + \varepsilon - \sin \varphi_{\text{diff}}}. \tag{S32}$$

The equation can be numerically integrated, and the results for $A = -1$ are shown in Fig.S3. When the detuning is close to the onset of synchronization, i.e. when $\varepsilon$ is sufficiently smaller than unity, the denominator of the integrand becomes small at its local minima and the integration requires a longer period of time for $\varphi_{\text{diff}}$ to change. This induces a partial stability in the phase synchronization wherein a quantized phase slip occurs. Clearly, the phase slip duration, $\tau$, becomes longer for a smaller detuning offset $\varepsilon$. This theoretical result explains the experimental results shown in Fig. 2B of the main text.

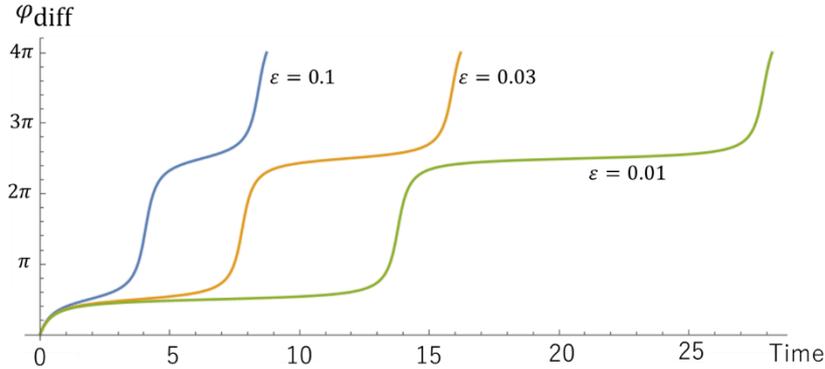

Fig. S3 Calculated time evolution of phase difference, $\varphi_{diff}$, for different detuning parameters, $\varepsilon$. The periodic phase slip and the slip interval, $\tau$, which becomes longer for smaller $\varepsilon$, are consistent with the experimental observations.

Next, we show the case of an overtone Kuramoto potential. The potential is given by $\zeta = \Delta\varphi_{diff} + A\cos(n\varphi_{diff})$. The integration (S28) for $\varepsilon = \frac{\Delta}{A} - 1$ is given by

$$t_f - t_i = -A^{-1} \int_{\varphi_i}^{\varphi_f} \frac{d\varphi_{diff}}{1 + \varepsilon - \sin(n\varphi_{diff})}. \tag{S33}$$

The minimum of the denominator appears with a period of $2\pi/n$. This is why the fractional phase slip $2\pi/n$ is observed in this case of an overtone potential. The results of a numerical integration of Eq. (S33) are shown in Fig. S4 for $A = -1$; they show good agreement with the experimental results in Fig. 2C of the main text.

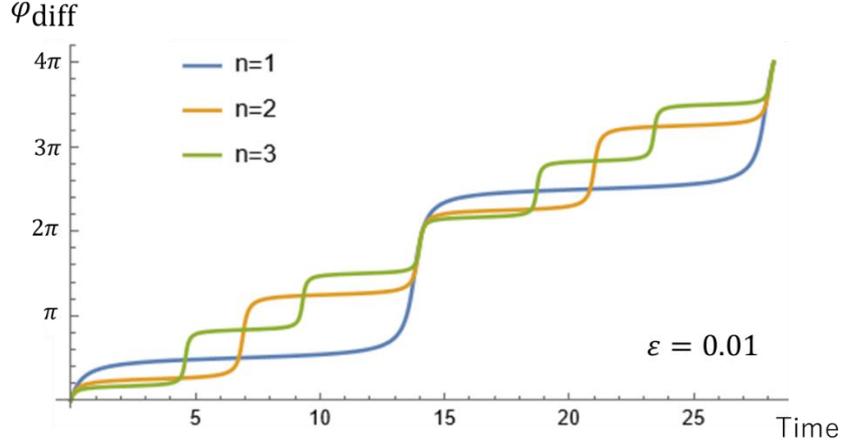

Fig. S4 Calculated time evolution of $\varphi_{diff}$ for overtone potentials having different orders ($n$). The fractional phase slip, $2\pi/n$, is in good agreement with experiments.

Next, we show an example of a synthesized overtone potential, $\zeta = \Delta\varphi_{diff} + A_1 \cos(\varphi_{diff} + \theta_1) + A_2 \cos 2\varphi_{diff}$. As in the previous examples, partial phase stability occurs when the force field $g(\varphi_{diff})$ becomes small. We assume that the local minimum of $g(\varphi_{diff})$ is always positive (or negative) for avoiding stable synchronization. The dynamics in this case strongly depend on $\Delta\varphi_{diff}$, $A_1$, $A_2$, and $\theta_1$ but, as an example, we will show the result for $\Delta\varphi_{diff} = -1.05$, $\theta_1 = \pi/2$ and $A_2 = 0.5$ in order to make a comparison with our experimental results.

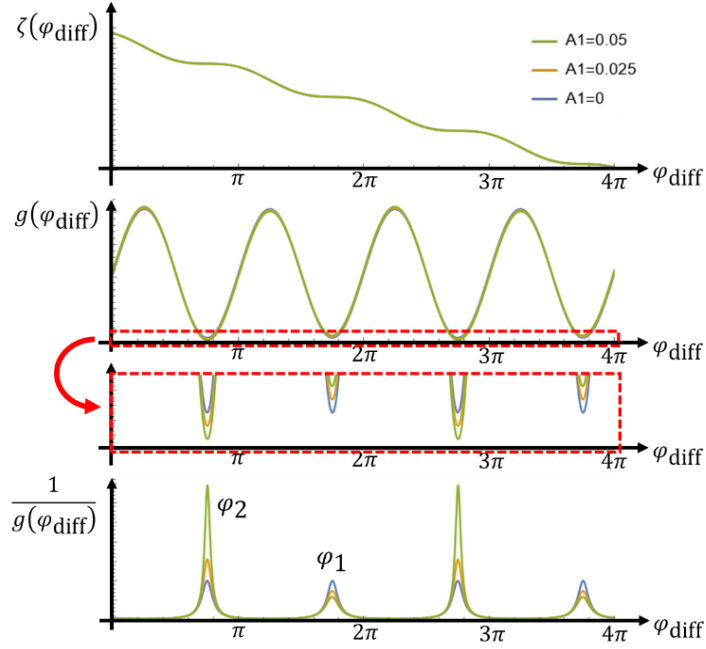

Fig. S5 $\zeta(\varphi_{\text{diff}})$, $g(\varphi_{\text{diff}})$, and the integrand $1/g(\varphi_{\text{diff}})$, for a synthesized overtone potential, $\zeta = \Delta\varphi_{\text{diff}} + A_1 \cos(\varphi_{\text{diff}} + \theta_1) + A_2 \cos 2\varphi_{\text{diff}}$, with $\Delta = -1.05$, $\theta_1 = \pi/2$ and $A_2 = 0.5$. Applying small modulation to the amplitude, $A_1$ induces a dramatic change in the integrand. The peak heights of the two partially stable phases ($\varphi_1$ and $\varphi_2$) are different, causing the asymmetric phase slip in Fig. 3 of the main text. The magnification of the red dashed-line box shows the plot in the vicinity of the horizontal axis.

Figure S5 shows $\zeta(\varphi_{\text{diff}})$, $g(\varphi_{\text{diff}})$, and the integrand $1/g(\varphi_{\text{diff}})$, for synthesized overtone potentials for different values of $A_1$ with $A_2$ set to 0.5. Applying a small modulation to the fundamental harmonic, $A_1 \cos(\varphi_{\text{diff}} + \theta_1)$, negligibly affects the potential $\zeta(\varphi_{\text{diff}})$ and the force field $g(\varphi_{\text{diff}})$, but it dramatically affects the integrand $1/g(\varphi_{\text{diff}})$. Slightly increasing $A_1$ induces a large difference in the peak height of $1/g(\varphi_{\text{diff}})$ between the even- and odd-numbered stable phases ($\varphi_1$ and $\varphi_2$), leading to an asymmetric double-period phase slip. This can be easily understood by inspecting the plot of $g(\varphi_{\text{diff}})$ near the horizontal axis (the magnification of the red dashed-line box in Fig. S5). The overall shapes of $\zeta(\varphi_{\text{diff}})$ and $g(\varphi_{\text{diff}})$ look unchanged, but the distance from the horizontal axis is drastically altered by the contribution from the fundamental harmonic, leading to a large change in the integrand. Figure S6 shows the calculated time evolution of $\varphi_{\text{diff}}$ for the three different values of $A_1$. When $A_1 = 0$, the dynamics are governed only by the second harmonic component, $A_2 \cos 2\varphi$, and phase slip of $\pi$ periodically occurs. However, increasing $A_1$ induces the asymmetry between the phase slip $n\pi \to (n+1)\pi$ and the phase slip $(n-1)\pi \to n\pi$, reproducing the experimental observations shown in Fig. 3 in the main text.

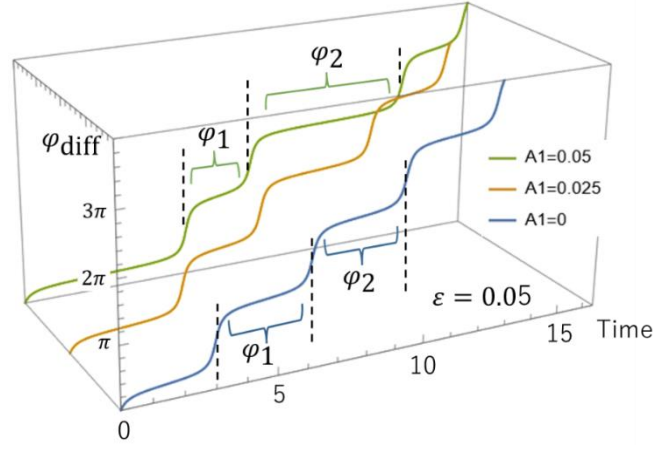

Fig. S6 Calculated time evolution of $\varphi_{\text{diff}}$ for different values of $A_1$. Even though $A_1$ is much smaller than $A_2 = 0.5$, the small modulation has a large effect of inducing an asymmetric phase slip, where two different slip durations alternately appear.

**Adiabatic modulation of the Kuramoto potential**

Here, we describe the phase dynamics in adiabatically modified Kuramoto potential. If we consider an adiabatic change of the Kuramoto potential, the phase dynamics is described as

$$\dot{\varphi}_{\text{diff}} = -\frac{\partial \zeta(\varphi_{\text{diff}}, t)}{\partial \varphi_{\text{diff}}}. \tag{S34}$$

Here, we explicitly indicate the time dependence of the potential as $\zeta(\varphi_{\text{diff}}, t)$. If the time dependence is sufficiently slow compared with the time required to stabilize the synchronization, we can apply the adiabatic approximation. As already discussed, the phase synchronization occurs at the phase difference where the Kuramoto potential $\zeta$ has a local minimum, i.e. the force field $g(\varphi_{\text{diff}}) = -\frac{\partial \zeta}{\partial \varphi_{\text{diff}}}$ vanishes with a negative slope. Then, we can assume that the synchronized $\varphi_{\text{diff}}$ adiabatically changes along the local minimum of $\zeta(\varphi_{\text{diff}}, t)$ following the time variation in the Kuramoto potential. Let's start from a simple example. Suppose we have the Kuramoto potential with a phase linearly shifted with time,

$$\zeta(\varphi_{\text{diff}}, t) = A \cos(\varphi_{\text{diff}} + 2\pi t/T). \tag{S35}$$

Moreover, we will assume no frequency detuning. For $A > 0$, the local minima of $\zeta$ is given by $\varphi_{\text{diff}} = (2n + 1)\pi - 2\pi t/T$. Therefore, if $T$ is large enough, the locked phase $\varphi_{\text{diff}}$ linearly shifts with time. This effect is in reality identical to the frequency detuning but the time-dependent phase shift is caused by the difference between the reference and modulation frequencies ($\Omega_{\text{diff}}$ and $\Omega_{\text{diff}}+2\pi/T$, respectively).

Next, let us consider a nontrivial example of an adiabatic potential change by introducing the time variation in the synthesized overtone potential. In this model, the phase of the fundamental harmonic is linearly shifted with time, whereas the second harmonic potential is time independent. The potential form is given by

$$\zeta(\varphi_{\text{diff}}, t) = 2k(1-r) \cos\left(\varphi_{\text{diff}} + \frac{2\pi t}{T}\right) + kr \cos(2\varphi_{\text{diff}}). \tag{S36}$$

In our experiments, the phase shift, $2\pi t/T$, was introduced by modifying the phase of modulation laser light, as described in main text. Here, $2k(1-r)$ and $kr$ correspond to $A_1$ and $A_2$, respectively. We have no detuning term so that a stable synchronization occurs, and no phase slip is induced if no adiabatic variation is introduced. However, we will show that the phase slip can be triggered by introducing the externally controlled phase shift, $2\pi t/T$, even within the synchronization condition. We will also show that the controlled phase slip utilizes the topological feature of time dependent Kuramoto potential. Such topological properties in periodically modulated one-dimensional systems, i.e., 1+1 dimensional systems, are analogous to the Thouless pump reported in cold atoms and photonic systems.

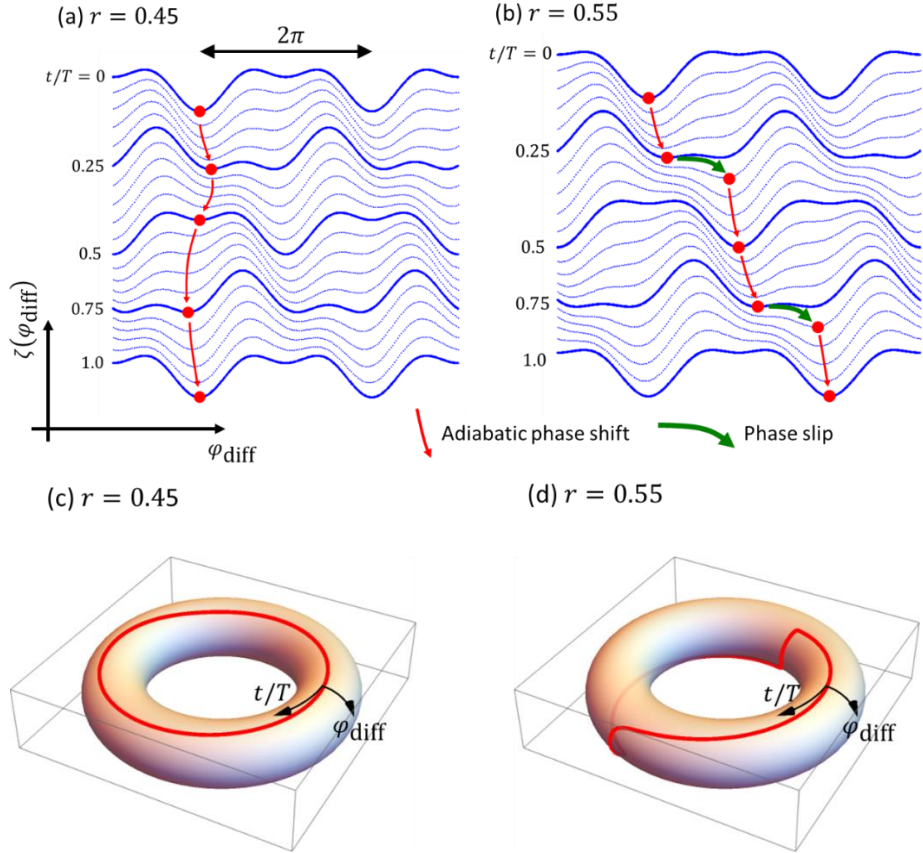

Fig. S7 Calculated time evolution for the Kuramoto potential given by (c11) (blue curves) and the variation of the local minimum (red circles) for (a) $k = -0.1, r = 0.45$ and (b) $k = -0.1, r = 0.55$. The phases of the two oscillators are synchronized at the local minimum and move adiabatically following the potential minimum. For $r = 0.45$, the local minimum continuously and periodically changes. In contrast, for $r = 0.55$, the local minimum disappears at $\varphi_{\text{diff}} = 0.25$ (and also at 0.75) and the phase slip is induced in the nearest neighbor local minimum. (c) and (d) show the phase trajectories mapped onto the torus surfaces. Different winding numbers $W$ are visible between $r = 0.45$ ($W = 0$) and $r = 0.55$ ($W = 1$).

Figure S7 (a) and (b) show the calculated time variation of the Kuramoto potential $\zeta(\varphi_{\text{diff}}, t)$ (blue curves) as well as that of the potential minimum (red circles) for $k = -0.1$. When $r = 0.45$ (a), the local minimum continuously and periodically changes within a small region and no discontinuous shift is observed. On the other hand, for $r = 0.55$ (b), the potential local minimum disappears when $t/T$ becomes 0.25 or 0.75 and a discontinuous phase slip (green arrows) is induced in the nearest-neighbor local minimum. Because the phase time variation is periodic with period $T$, we can map the phase trajectories onto torus surfaces (c) and (d). It is clear that the two trajectories show different winding numbers, where $r = 0.45$ shows no winding, whereas $r = 0.55$ shows a unit winding. The transition between these two topologically different trajectories occurs at $r = 0.5$. These results indicate that the phase dynamics exhibit a discontinuous transition at $r = 0.5$ reflecting the topological features of the 1+1 dimensional Kuramoto potential.

**Hysteresis and non-reciprocal dynamics**

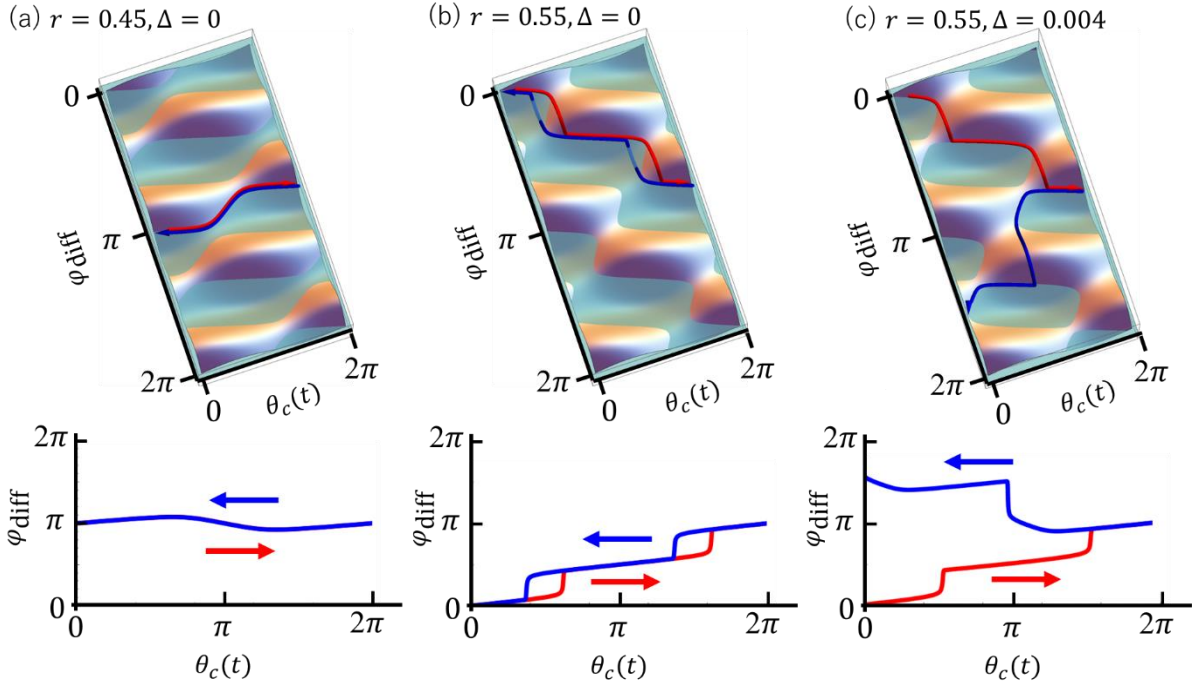

Fig. S8 Top three panels showing the calculated phase trajectories in $\varphi_{\text{diff}}$ - $\theta_c$ 1+1 phase space for both upward (red curves) and downward (blue curves) sweeps. Superimposed on the plots are a bird's-eye view of the force field ($g(\varphi_{\text{diff}}, \theta_c)$) and a horizontal plane $g(\varphi_{\text{diff}}) = 0$, by which the negative $g(\varphi_{\text{diff}}, \theta_c)$ region is colored light green and partially transparent. Bottom panels showing the calculated $\varphi_{\text{diff}}$ when $\theta_c$ is swept from 0 to $2\pi$ (red) and is swept back from $2\pi$ to 0 (blue). The numerical calculations used the parameters, $k = -0.1$ and (a) $r = 0.45, \Delta = 0$, (b) $r = 0.55, \Delta = 0$, and (c) $r = 0.55, \Delta = 0.004$.

Here, we describe the hysteresis and non-reciprocal dynamics observed with the adiabatically modulated Kuramoto potential. For this purpose, it is convenient to examine the phase trajectory in 1+1-dimensional phase space. We again use a Kuramoto potential similar to (S36) but with the detuning $\Delta$ and an arbitrary time dependence to the phase shift, $\theta_c(t)$,

$$\zeta(\varphi_{\text{diff}}, \theta_c(t)) = \Delta \varphi_{\text{diff}} + 2k(1 - r)\cos(\varphi_{\text{diff}} + \theta_c(t)) \quad (S37)$$
$$+ kr\cos(2\varphi_{\text{diff}}).$$

In articular, we consider the case that the additional phase $\theta_c(t)$ takes an adiabatic round-trip between $\theta_1 = 0$ and $\theta_1 = 2\pi$ as a function of $t$. The top panels in Fig. S8 show the calculated phase trajectories for $k = -0.1$ together with a superimposed bird's-eye view of the force field $g(\varphi_{\text{diff}}, \theta_c) = -\partial \zeta(\varphi_{\text{diff}}, \theta_c)/\partial \varphi_{\text{diff}}$. As already discussed, the phase trajectory traces the local minimum of $\zeta(\varphi_{\text{diff}}, \theta_c)$, i.e. the zero-value curve of the force field, $g(\varphi_{\text{diff}}, \theta_c) = 0$, with a negative slope, $\frac{\partial g(\varphi_{\text{diff}}, \theta_c)}{\partial \varphi_{\text{diff}}} < 0$. To visualize the zero-value curves, the horizontal plane $g(\varphi_{\text{diff}}, \theta_c) = 0$ is also plotted and the negative $g(\varphi_{\text{diff}}, \theta_c)$ regions are colored light green and partially transparent. The bottom panels in Fig. S8 show the change in the stable $\varphi_{\text{diff}}$ when $\theta_1$ is swept from 0 to $2\pi$ (red curves) and swept back from $2\pi$ to 0 (blue curves).

As already described in the case of $\Delta = 0$, no phase slip appears for (a) $r < 0.5$, whereas a discontinuous phase slip occurs when (b) $r > 0.5$. It is interesting to see the backward sweep, where clear hysteresis is observed only for $r > 0.5$. The landscape of the force field clearly shows the origin of hysteresis. The discontinuous slip occurs when the stable zero-value point is terminated in the $\theta_c$ direction, and the slip position is different between the forward (red curve) and backward (blue curve) sweeps. In addition, the topological features of the potential landscape are different between (a) and (b). The $2\pi$ forward rotation of $\theta_c$ leads to a similar forward rotation of $\varphi_{\text{diff}}$ in (b), but not in (a), inducing different winding numbers of 1 and 0, respectively. An even more interesting feature is observed when the detuning Δ is finite (c). In this case, the zero-value curve forms a closed loop in 1+1 phase space, and the phase slip induces a unidirectional average drift in the positive $\varphi_{\text{diff}}$ direction. These highly nonreciprocal dynamics and hysteresis are clearly caused by the topological features of the Kuramoto potential, which caused the various winding numbers and nonreciprocal dynamics observed in our experiments.

**Basic experimental setup**

An optomechanical microbottle resonator was fabricated on a silica glass fiber whose diameter was 80 μm via the heat and pull technique (S4). It had a maximum bottle diameter of 80 μm, neck diameter of 78 μm, and a separation length between the two necks of about 500 μm (see Fig. S9).

To perform synchronization experiments, an external cavity diode laser (ECDL) was used to excite the optomechanical coupling by controlling its intensity and polarization with an erbium-doped fiber amplifier (EDFA) and polarization controller (PC). To dynamically modulate the input laser intensity, an optical intensity modulator was utilized with radio-frequency modulation signals generated by an arbitrary function generator (AFG). The input laser light was coupled to an optical whispering-gallery mode (WGM) with a silica tapered fiber which touched the microbottle resonator. The output light from the resonator was detected by a photodetector whose dc part, i.e., optical transmission, was measured with a digital signal oscilloscope (DSO), and ac part, i.e., mechanical vibration signals, was measured with an electronic spectrum analyzer (ESA). Moreover, the beat signal between two mechanical oscillations was detected by filtering out the two oscillating signals around 48 MHz and measuring the self-mixed signals with a lock-in amplifier. A schematic diagram of the setup is shown in Fig. S9.

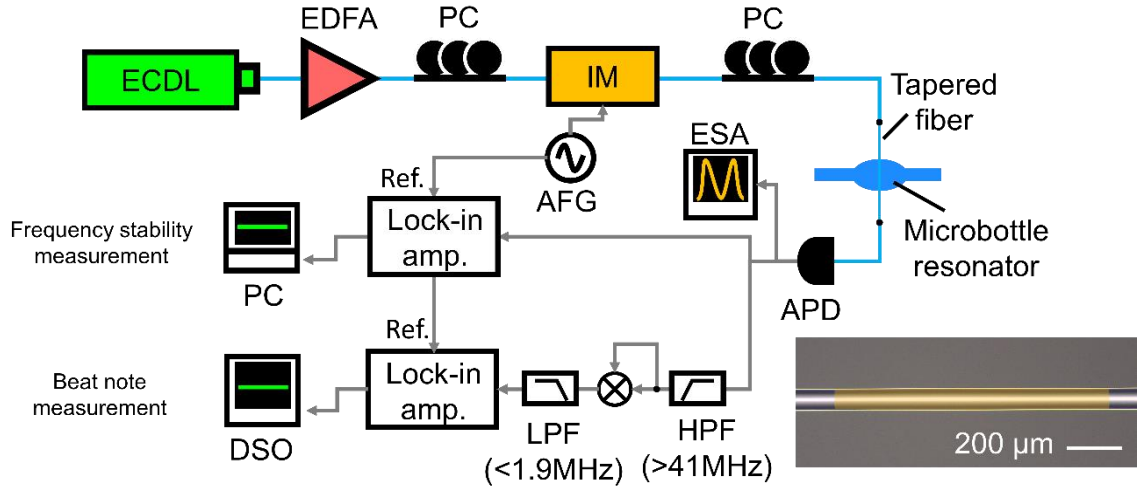

S9. Schematic experimental setup and optical microscope image of the silica microbottle resonator. The orange shaded area in the microscope image corresponds to the microbottle structure.

**Synchronization with over-tone modulation**

The synchronization conditions were also investigated in over-tone modulations $\Omega_{\text{mod}} \sim n\Omega_{\text{diff}}$ for $n=2$, 3, and 4. Fig. S10 shows color maps of the measured amplitude of the two-mode beat notes. As in Fig. 1E, the higher beat-note amplitude reflects phase synchronization because the two-mode oscillations are mode locked. These figures confirm that integer multiples of the modulation frequency also lead to higher-order synchronization.

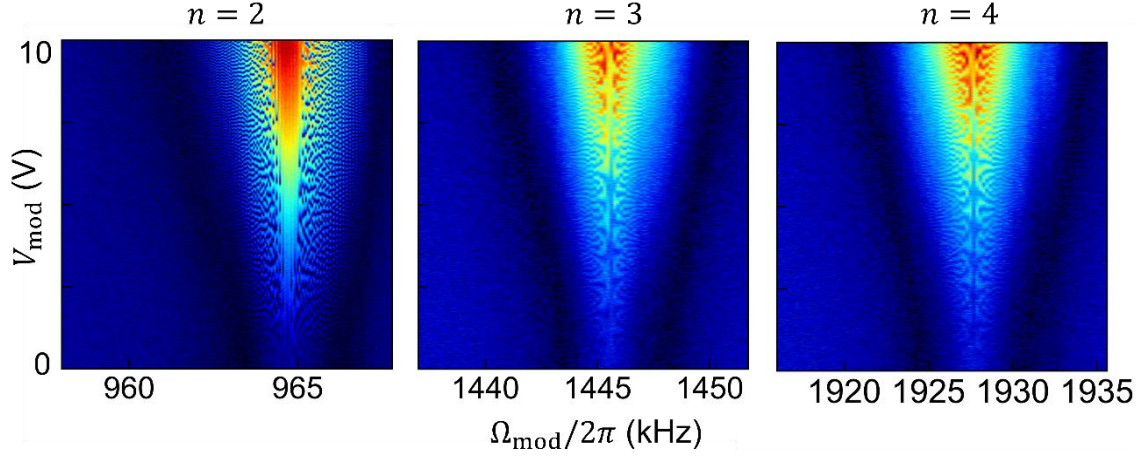

S10. Arnold tongues for higher-order synchronizations.

**Adiabatic amplitude modulation technique**

To precisely explore synchronization condition, the modulation depth of the signals from the AFG $s(t)$ was linearly swept as

$$s(t) = V_{\text{mod}}(1 - t/T_{\text{ad}}) \cos \Omega_{\text{mod}} t, \qquad (S38)$$

where $T_{\text{ad}}$ is longer than any physical characteristic time such as dissipation, coupling, and oscillation to be an adiabatic formation of synchronization potential. Here, we set $T_{\text{ad}}$ to a typical value, 0.25 sec.

For the measurement of the 1:1 synchronization in Fig. 1E, we measured the beat note between the two mechanical oscillation signals. Note that the crosstalk from the optical modulation itself around the frequency of $\Omega_{\text{mod}}$ was removed by the HPF that picked up the two mechanical oscillation signals around $\Omega_M \sim 48$ MHz. Accordingly, the intensity of the self-mixed beat signal reflects how well locked the two modes are. Thus, the beat signals along the vertical axes in Fig. 1E were measured in a fast sweep with the adiabatic amplitude modulation technique. This experimentally revealed the synchronization condition in the form of the clear Arnold's tongue.

This sweeping technique of the potential modulation was also used to extract the scaling of phase slip duration, $\tau$, shown in Fig. 2D. In this scheme, we set the finite detuning $\Delta$ and swept the signal to be $s(t) = V_{\text{max}}(1 - \eta t/T_{\text{ad}}) \cos \Omega_{\text{mod}} t$, where $V_C$ was the critical voltage at which the phase slip began to be observed in experiment, and $\eta = V_{\text{min}}/V_{\text{max}}$ determines the sweeping window. Fig. S11 shows a conceptual illustration and example data from which the scaling law of the phase slip was derived with the adiabatic amplitude modulation technique.

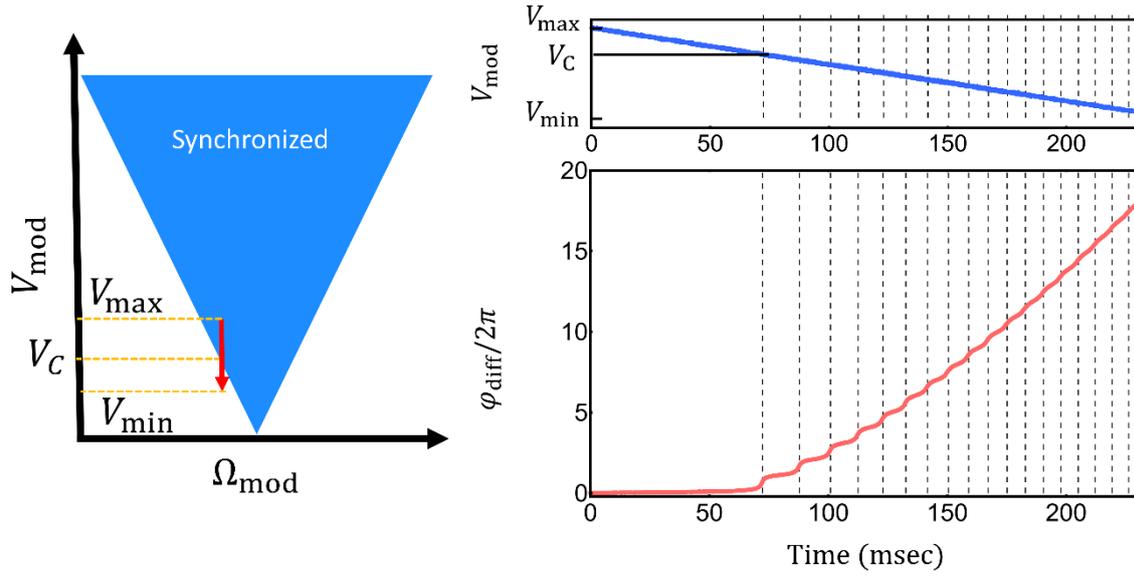

S11. Conceptual illustration of adiabatic amplitude modulation technique to observe the scaling law in the phase slip (left) and example of data obtained in the experiment (right).

**Theoretical analysis of topological path in dynamical synchronization**

We theoretically investigated the total phase slip in the toroidal phase space by numerically solving the equation of motion of the phase difference, $\varphi_{\text{diff}}$, in the Kuramoto model,

$$\dot{\varphi}_{\text{diff}}(t) = \Delta + A_1 \sin(\varphi_{\text{diff}} + \theta_C(t)) + A_m \sin m\varphi_{\text{diff}}, \quad (S39)$$

where $\Delta$ is the detuning, $A_j$ is the depth of the potential with respect to the integer $j$, $\theta_C(t) = 2\pi t/T$ is the linearly modulated potential phase with the adiabatic period $T$. In the numerical calculation, we first found the initial equilibrium points $x_j(0)$ ($j = 1, 2, \ldots, m$), and then numerically solved Eq. (S39) to achieve the trajectory $x_j(t)$ for each initial equilibrium point from $t = 0$ to $t = T$. We theoretically reproduced the experimental results, as shown in Fig. S12. By carefully choosing the parameters under $A_m = 1$, we obtained good correspondence to the experimentally obtained path, including the fractional winding number [S11(B), (D), (F), and (H)].

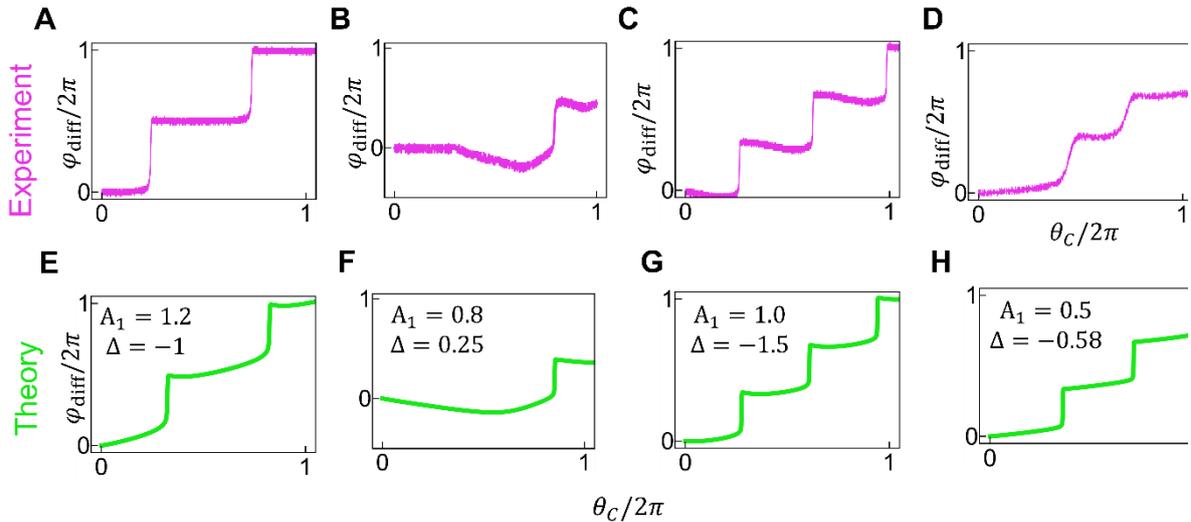

S12. Topological path under dynamical control of synchronization potential observed in experiment (A)-(D) and numerically calculated (E)-(H).

S1. C. C. Rodrigues, C. M. Kersul, A. G. Primo, M. Lipson, T. P. M. Alegre, and G. S. Wiederhecker, Optomechanical synchronization across multi-octave frequency spans. *Nat. Commun.* 12, 5625 (2021).
S2. F. Dorfler and F. Bullo, Synchronization in complex networks of phase oscillators: A survey, *Automatica* 50, 1539–1564 (2014).
S3. P. Bhansali and J. Roychowdhury, Gen-Adler: The generalized Adler's equation for injection locking analysis in oscillators, Proceedings of the 14th Asia South Pacific Design Automation Conference 2009, pp. 522-527.
S4. S. Spillane, T. Kippenberg, O. Painter, and K. Vahala, Ideality in a fiber-taper-coupled microresonator system for application to cavity quantum electrodynamics, *Phys. Rev. Lett.* 91, 043902 (2003).